\documentclass[11pt,a4paper]{article}
\pdfoutput=1
\usepackage{jheppub}
\usepackage[T1]{fontenc} 

\usepackage{epsf,epsfig,subfigure,graphicx,amsmath,amssymb}
\usepackage{amsfonts}
\usepackage{latexsym}
\usepackage{color}
\usepackage{accents}
\usepackage{dhucs}
\usepackage{hyperref}
\usepackage{verbatim}
\usepackage{slashed}
\usepackage{notes2bib}
\usepackage{makecell}

\def\bea{\begin{eqnarray}}
\def\eea{\end{eqnarray}}

\def\be{\begin{equation}}
\def\ee{\end{equation}}

\def\Mp{M_{\rm pl}}
\def\Mpl{M_{\rm pl}}

\def\TeV{\,{\rm TeV}}
\def\GeV{\,{\rm GeV}}
\def\MeV{\,{\rm MeV}}
\def\keV{\,{\rm keV}}
\def\eV{\,{\rm eV}}

\def\eps{\epsilon}

\def\vew{v_{\textrm {\tiny EW}}}
\def\mbr{\mu_{\rm b}}

\def\Lbr{\Lambda_{\rm br}}

\title{\center{Probing the Relaxed Relaxion\\ at the Luminosity and Precision Frontiers}}
\author[a]{Abhishek Banerjee,}
\author[a]{Hyungjin Kim,}
\author[a]{Oleksii Matsedonskyi,}
\author[a]{Gilad Perez}
\author[b,c]{and Marianna S. Safronova}

\affiliation[a]{Department of Particle Physics and Astrophysics,\\ Weizmann Institute of Science, Rehovot 7610001, Israel}
\affiliation[b]{Department of Physics and Astronomy, University of Delaware, Newark, Delaware 19716, USA}
\affiliation[c]{Joint Quantum Institute, National Institute of Standards and Technology and the University of Maryland, Gaithersburg, Maryland 20742, USA}

\emailAdd{abhishek.banerjee@weizmann.ac.il}
\emailAdd{hyungjin.kim@weizmann.ac.il}
\emailAdd{oleksii.matsedonskyi@weizmann.ac.il}
\emailAdd{gilad.perez@weizmann.ac.il}
\emailAdd{msafrono@udel.edu}

\abstract{
Cosmological relaxation of the electroweak scale is an attractive scenario addressing the gauge hierarchy problem. Its main actor, the relaxion, is a light spin-zero field which 
dynamically relaxes the Higgs mass with respect to its natural large value.   
We show that the relaxion is generically stabilized at a special position in the field space, which leads to suppression of its mass and potentially unnatural values for the model's effective low-energy couplings. In particular, we find that the relaxion mixing with the Higgs can be several orders of magnitude above its naive naturalness bound. 
Low energy observers may thus find the relaxion theory being fine-tuned although the relaxion scenario itself is constructed in a technically natural way. 
More generally, we identify the lower and upper bounds on the mixing angle.
We examine the experimental implications of the above observations at the luminosity and precision frontiers. 
A particular attention is given to the impressive ability of future nuclear clocks to search for rapidly oscillating scalar ultra-light dark matter, where the future projected sensitivity is presented. 
}

\begin{document}
\maketitle
\flushbottom

\section{Introduction}\label{sec:intro}
The idea that the electroweak (EW) scale is determined as a result of dynamical relaxation provides a new insight on hierarchy problem in the Standard model (SM)~\cite{Graham:2015cka}. 
The electroweak scale is dynamically selected by an evolution of an axion-like field $\phi$, which is referred to as \emph{relaxion}.
Compared to conventional models for the electroweak hierarchy problem, the relaxation mechanism includes one infrared degree of freedom, relaxion, an axion like particle (ALP) which couples feebly to SM particles via its mixing with Higgs boson due to presence of CP violation~\cite{Choi:2016luu, Flacke:2016szy,Davidi:2017gir}.
The relaxion phenomenology is generally different from that of conventional solutions to the hierarchy problem, which exploits high-energy colliders that focus on the TeV scale.
Experimental programs searching for weakly coupled light scalar and pseudo-scalar particles are thus more suitable for relaxion searches~\cite{Kobayashi:2016bue, Choi:2016luu, Flacke:2016szy, Frugiuele:2018coc, Budnik:2019olh}.

Minimal relaxion models were shown to lead to a viable ALP dark matter (DM) candidate~\cite{Banerjee:2018xmn}, which gives rise to a variety of interesting signals associated with the fact that effectively the Higgs vacuum expectation value (VEV) oscillates with time. The corresponding signals were discussed in the context of dilaton DM~\cite{Arvanitaki:2014faa}. However, the relaxion DM model seems to prefer rapidly oscillating frequencies which pose both challenging and exciting~\cite{Aharony:2019iad,  Safronova:2017xyt,PhysRevResearch.1.033187, Antypas:2019yvv} quest for variety of precision-front experiments.

The relaxion potential consists of two parts: one for scanning the Higgs mass, and the other one for providing a feedback to the relaxion evolution as a function of the Higgs VEV. 
Specifically, we consider 
\bea
V(\phi,H) = 
\mu_H^2(\phi) |H|^2 
+ \lambda |H|^4
+ V_{\rm roll}(\phi)
+ V_{\rm br}(\phi,H)\, ,
\label{rel_potential}
\eea
where
\bea
\mu_H^2(\phi) &=& \Lambda^2 - g \Lambda \phi\,,
\label{eq:muh}
\\
V_{\rm roll}(\phi) &=& - g \Lambda^3\phi\,,
\label{eq:vroll}
\\
V_{\rm br}(\phi,H) &=& -\mbr^2 |H|^2 \cos (\phi/f)\,.
\label{br}
\eea
Here, $\Lambda$ is the Higgs mass cutoff scale and $\mbr\lesssim \vew$ is the backreaction scale with $\vew= 174\,$GeV.\footnote{See~\cite{Espinosa:2015eda} for possible generalizations of the backreaction potential.}
We consider the relaxation scenario during the primordial inflation (see also~\cite{Hook:2016mqo, Fonseca:2018xzp, Fonseca:2019lmc} for the relaxation mechanism based on particle production).
At the very beginning of the relaxion evolution, the Higgs mass takes a large positive value, which is of order the cutoff scale. 
Relaxion rolls down its potential $V_{\rm roll}$ \eqref{eq:vroll}, and the Higgs mass~(\ref{eq:muh}) decreases continuously. 
At a critical field value, $\phi_c = \Lambda / g$, the Higgs mass changes its sign, and nonvanishing vacuum expectation value is developed. 
As a consequence, the backreaction potential $V_{\rm br}$ is generated, providing a periodic potential barrier. 
The relaxion finds the classically stable minimum when the slope of the rolling potential balances with that of the backreaction potential, 
\be\label{eq:stop}
V_{\rm br}' \simeq - V'_{\rm roll}\,.
\ee
The relaxion parameters can be engineered in a natural way that this condition is met only when $\langle H \rangle = \vew$, that is $g\Lambda^3f = \mbr^2 \vew^2$, and hence, the electroweak scale is dynamically chosen by the relaxion field.\footnote{For that to happen, the relaxion mechanism generally requires a large field excursion, $\Delta \phi / f \sim (\Lambda^4/ \mbr^2 \vew^2) \gg 1$, but this can be  achieved in a technically natural way by clockwork mechanism~\cite{Choi:2014rja, Choi:2015fiu, Kaplan:2015fuy, Giudice:2016yja}.}
Assuming that $\cos(\phi_0/f) \sim \sin(\phi_0/f) \sim 1$ at the minimum of the potential $\phi_0$, the mass of relaxion would be naively given as 
\be\label{eq:mphinaive1}
(m_{\phi }^{2})_{\rm naive} \,\sim\, \partial^2_\phi \, V_{\rm br}(\phi,H) \,\sim\,  \mbr^2 \vew^2/f^2\,,
\ee 
while the mixing angle with the Higgs would be given as 
\be\label{eq:naivemixing}
(\sin\theta_{h\phi})_{\rm naive} 
\,\sim\,  \partial_\phi \partial_h \, V_{\rm br}(\phi,H) /m_h^2 
\,\sim\,  \mbr^2 / \vew f \, .
\ee 
The combination of the mass and the mixing angle has a paramount importance for the experimental detection of the relaxion throughout its all parameter space. Another parameter that might play an important role in the relaxion phenomenology for $m_{\phi }\gtrsim\,\MeV$ is the relaxion-Higgs quartic coupling, $\lambda_{H\phi} \,,$ whose naive estimate is 
\be\label{eq:naivequartic}
(\lambda_{H\phi})_{\rm naive} \,\sim\,  \partial^2_\phi \partial^2_h \, V_{\rm br}(\phi,H)  \,\sim\,  \mbr^2 /f^2\,\sim\,(m_{\phi }^{2})_{\rm naive}/\vew^2\,\,.
\ee 

In this paper, we show that the naive estimates~\eqref{eq:mphinaive1},~\eqref{eq:naivemixing},~\eqref{eq:naivequartic} and the relations among them could be violated by several orders of magnitude, and analyse the implications of this fact for the relaxion experimental detection.

\section{The relaxion mass and couplings: a closer look}
\begin{figure}[t]
\centering
\includegraphics[scale=0.5]{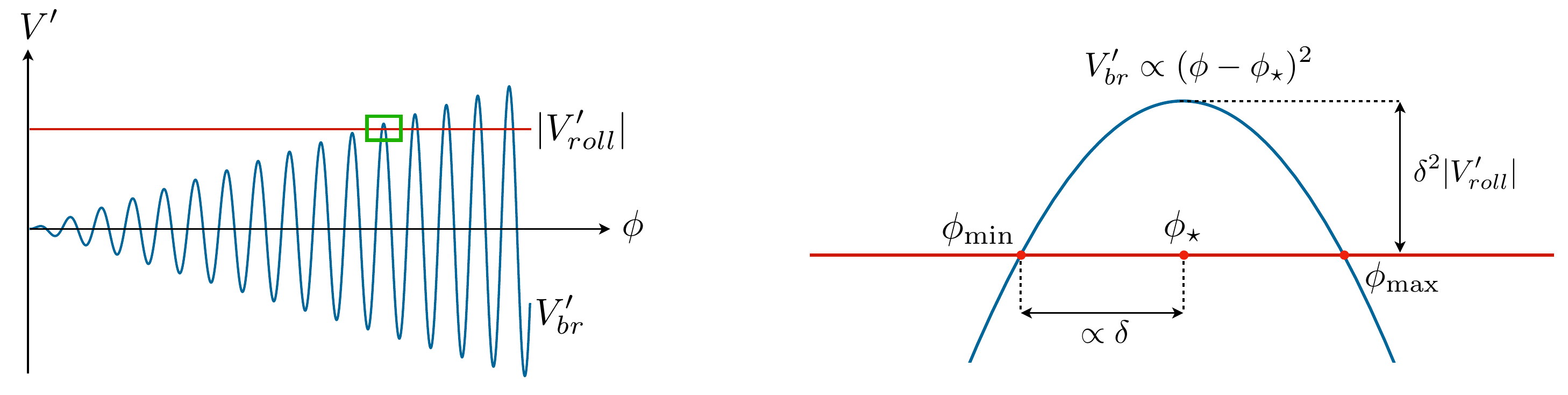}
\caption{Schematic picture of the slope of relaxion potential in the neighbourhood of the first local minimum (left panel), and zoomed in around the first minimum (right panel), with the Higgs field set to the minimum of its potential. The first relaxion minimum is reached when the slope of the rolling potential (red line) is balanced by the slope of the periodic barriers (blue line). $\phi_{\rm min}$, $\phi_{\rm max}$, and $\phi_\star$ are the positions of the first minimum, the first maximum, and the  inflection point of $V$, respectively. }
\label{fig:sketch}
\end{figure}

The above simple and rather naive way to estimate the relaxion mass and couplings requires a more careful look. Here we briefly summarize our main results, while we discuss detailed derivations and their implications in later sections.
As was already noticed in the original paper~\cite{Graham:2015cka},  
the relaxion mass is naturally suppressed compared to~\eqref{eq:mphinaive1} at the first minimum. 
The reason is that the amplitude of the backreaction potential grows only incrementally. For every oscillation period, $\Delta \phi = 2 \pi f$, the fractional change of the Higgs mass is $\mbr^2/\Lambda^2$, and so is the change of the amplitude of the backreaction potential~\eqref{br}. 
Since two small numbers appear repeatedly in the following discussion, we define them here as
\bea
\delta &\equiv& \frac{\mbr}{\Lambda}\,,
\label{d}
\\
\eps &\equiv& \Big(\frac{\mbr}{\vew}\Big)^2,
\label{e}
\eea
where $\delta$ is related to an incremental change of the Higgs mass over one period, and $\eps$ is the sensitivity of the Higgs mass to the relaxion field, $\eps \sim \partial \ln v^2 / \partial (\phi/f)$. 
Due to the incremental change of the Higgs mass, after the first minimum is reached, the backreaction potential barely grows above the value needed to compensate the linear slope -- at most by a $\delta^2$ fraction -- and quickly drops. This behaviour is schematically shown in the left panel of Fig.~\ref{fig:sketch}. Given that the shape of $V^\prime_{\rm br}$ around its maximum $\phi_\star$ is well approximated by the quadratic potential, $-(\phi - \phi_\star)^2$, and an overall change of $V^\prime_{\rm br}$ between $\phi_\star$ and the first minimum $\phi_{\rm min}$ is $\propto \delta^2$, it follows that $(\phi_{\rm min} - \phi_\star) \propto \delta$~\cite{Banerjee:2018xmn}. This is shown graphically in the right panel of Fig.~\ref{fig:sketch}. Moreover, for the relaxion mass, we obtain $m_\phi^2 = V_{\rm br}''(\phi_{\rm min})  \propto (\phi_{\rm min} - \phi_\star) \propto \delta$. In other words, the vicinity to the inflection point $\phi_\star$ suppresses the $\phi$ mass such that (see Section~\ref{sec:relaxation})
\bea
m_\phi^2 \simeq \frac{\mbr^2 \vew^2}{f^2} \frac{\mbr}{\Lambda} \simeq (m_{\phi }^{2})_{\rm naive}\times \delta\, ,
\eea
from which one sees that the relaxion mass contains additional suppression factor of $\delta$  with respect to the naive estimate~(\ref{eq:mphinaive1}). At the same time, the potential barrier height at the first local minimum is suppressed by $\delta^3$ relative to $\mbr^2 \vew^2$ (see Section~\ref{sec:relaxation}).  
This is important for the stability of this minimum during inflationary epoch. If the first minimum is not stable and the $n$-th local minimum is populated instead, the relaxion mass would also change accordingly.
Such cosmological evolution will be further discussed in Section~\ref{sec:relaxation}. Given the repeated appearance of some parameters with different subscripts through out the text, such as $m_{\phi}$ and $\sin\theta_{h\phi}$, we have presented a glossary in Table~\ref{tab:glossary} to describe their differences.

Another interesting aspect of dynamical relaxation is the fact that the mixing angle with the Higgs boson is bounded both from above and below for a given mass, forming a compact two-dimensional relaxion parameter space. An important feature defining this parameter space is the fact that the relaxion mass is suppressed by additional small parameter $\mbr/\Lambda$, whereas the mixing angle is given as one naively expected.
For a broad range of relaxion mass, we find the maximum mixing angle (see Section~\ref{sec:maxmix})
\bea
(\sin\theta_{h\phi} )_{\rm max} \simeq \left( \frac{m_\phi}{\vew} \right)^{2/3}\!.
\label{max_mix}
\eea
This bound has important implications for the experimental testability of this scenario, but it is also important in the context of naturalness arguments. 
Low energy observers may find this theory fine-tuned since the cosmological relaxation scenario allows relaxion mixing angle and mass such that the radiative correction to the mass could be larger than the tree level contribution.
To see this point, suppose that one has measured the relaxion mass and mixing angle corresponding to the latter maximum value, given in Eq.~\eqref{max_mix}, so that the relative size of the quantum correction is (see Section~\ref{sec:low_obs})
\bea
\frac{\Delta m_\phi^2}{m_\phi^2} \sim  \frac{(\sin\theta_{h\phi} )^2_{\rm max}}{16\pi^2} \frac{\vew^2}{m_\phi^2} \sim \frac{1}{16\pi^2} \left( \frac{\vew}{m_\phi} \right)^{2/3}\!.
\eea
The above ratio could greatly exceed unity in a wide range of viable relaxion masses. In this case, an observer may conclude that what she/he has measured is radiatively unstable, and hence fine-tuned, although the relaxion model itself is constructed in a technically natural way. This conclusion results from the violation of the natural mass-mixing angle relation
~\cite{Piazza:2010ye, Arvanitaki:2015iga, Graham:2015ifn},
\bea
(\sin\theta_{h\phi} )_{\rm nat} \lesssim \frac{ m_\phi}{\vew}\, ,
\label{nat_mix}
\eea
which is stronger than the upper bound of Eq.~\eqref{max_mix}. 

In some sense, not only the Higgs mass is relaxed from its cutoff scale to the EW scale, but also \emph{the relaxion itself is relaxed, opening up the parameter space to the point which could be thought unnatural in view of conventional naturalness argument.}
We investigate the maximum and minimum mixing angle as well as the conventional naturalness argument in the context of relaxion scenario in Section~\ref{sec:low_obs} and in Section~\ref{sec:mixing_angle}. 
Similar discussion also holds to the quartic coupling, $\lambda_{H\phi}\,$, which, as discussed in Section~\ref{sec:xquartic}, can violate the corresponding naturalness criterion $\lambda_{H\phi}\lesssim m_\phi^2/\vew^2\,$.

Having determined parametric dependences of the relaxion mass and mixing angle on model parameters, we update the relaxion phenomenology in Section~\ref{sec:mixing_angle}. 
In addition to previously studied constraints on relaxion parameter space, we also discuss the reach of atomic physics probes at the low mass range of relaxion parameter space, assuming that the relaxion constitutes dark matter in the present universe.
Since the coherent oscillation of relaxion leads to an oscillation of fundamental constants, atomic tabletop experiments could provide efficient ways to probe the parameter space of an ultralight relaxion dark matter. Below we discuss the available constraints from currently available atomic clock systems and provide some projection for the sensitivity of future nuclear clock.

\begin{table}[t]
\centering
\begin{tabular}{ l  l  l }	
Notation
&  
&
\\ \hline \hline

$(m_\phi)_{\rm naive}$
& Relaxion mass when $\sin\theta_0 \sim \cos \theta_0 \sim 1$
&Eq.~\eqref{eq:mphinaive1}

\\ \hline

$(\sin\theta_{h\phi})_{\rm naive}$
& Relaxion-Higgs mixing angle when $\sin\theta_0 \sim \cos \theta_0 \sim 1$
&Eq.~\eqref{eq:naivemixing}
		
\\ \hline\hline

$m_{\phi 0}$
& Bare relaxion mass [$m_{\phi 0}^2= (m_\phi)^2_{\rm naive}\cos \theta_0$]
&Eq.~\eqref{mphi0}

\\ \hline
			
$m_{\phi}$ 
& Physical Relaxion mass 
&Eq.~\eqref{eq:relaxedmass} 

\\ \hline
$(\sin\theta_{h\phi})_{\rm max}$
& Maximum relaxion-Higgs mixing angle 
& Eqs.~\eqref{eq:mixmax1}--\eqref{eq:mixmax3}

\\ \hline
$(\sin\theta_{h\phi})_{\rm min}$
& Minimum relaxion-Higgs mixing angle 
& Eq.~\eqref{eq:mixmin1}

\\ \hline
\end{tabular}		
\caption{Glossary. Here $\theta_0 = \phi_0/f$ denotes the field value at a local minimum. The parameters with subscript ``naive'' indicate the quantity when $\sin\theta_0\sim\cos\theta_0 \sim1$. For more information, see the discussions around the quoted equations. }
\label{tab:glossary} 
\end{table}

\section{Relaxation of relaxion at $\pi/2$}\label{sec:relaxation}
\subsection{Vacuum structure}

We provide a heuristic argument why the relaxion mass is parametrically suppressed by additional small parameter $\delta$, while leaving more rigorous discussions in the Appendix~\ref{app:detail_pi/2} for interested readers.
For the backreaction potential under the consideration~\eqref{br}, this parametric suppression is closely related to the fact that the relaxion stops around the inflection point, in this case, $(\phi /f) \, {\rm mod} \, 2 \pi \simeq \pi/2$. 
As we will see below, this parametric suppression is not limited to the backreaction potential~\eqref{br}, but is a generic feature of any periodic potential as long as the amplitude of the backreaction potential changes incrementally, controlled by a small parameter $\delta$.

We focus on the relaxion evolution just before it eventually finds the electroweak scale Higgs mass.
For the following discussion, we rewrite the potential in terms of dimensionless angle parameter $\phi /f = 2 \pi m + \theta$ with $m \in {\mathbb Z}$ and $\theta \in [0,2\pi)$.
The minimum can be found by solving two equations 
\bea
\frac{\partial V}{\partial H} = 0\,,
\quad
\frac{\partial V}{\partial \phi} = 0\,. 
\label{eq_min}
\eea
The solution to the first equation can be straightforwardly found as
\bea
\langle |H|^2 \rangle \equiv v_m^2(\theta) = \frac{1}{2} \left[ -\mu_{H,m}^2(\theta) + \mbr^2 \cos\theta \right]\,,
\label{Higgs_VEV}
\eea
with $\mu_{H,m}^2(\theta) = \Lambda^2 - g \Lambda f ( 2\pi m + \theta)$. 
We assume that the Higgs quartic is $\lambda =1$ for brevity. 
Substituting this relaxion-dependent Higgs VEV to the potential, we define effective relaxion potential,
\bea
V_{\rm eff}(\theta) = V(\theta, \langle |H|^2 \rangle) = - \Lbr^4 \theta - v_m^4(\theta)\,,
\eea
where we have omitted unimportant constant term in the potential, defined $\Lbr^4 \equiv \mbr^2 \vew^2 = g\Lambda^3 f$,\footnote{ A precise condition that ensures $\langle |H|^2 \rangle = \vew^2$ is $g\Lambda^3 f = \mbr^2 \vew^2 \sin\theta_0 (1+\vew^2/\Lambda^2)^{-1}$. 
However, the difference between the correct condition and $g\Lambda^3 f = \mbr^2 \vew^2$ only gives subdominant corrections to the final results. } and the vacuum expectation value of the Higgs as $\vew = 174 \GeV$. 
The first term is a usual rolling potential, while the second term is the effective backreaction potential at the instantaneous minimum along the Higgs direction. 
For the notational convenience, we define the effective backreaction potential as
\bea
\widetilde{V}_{\rm br} \equiv - v_m^4(\theta)\,. 
\eea

The first relaxion minimum can be found by solving $\partial V_{\rm eff} / \partial \theta = 0$. 
It is equivalent to solve Eq.~\eqref{eq_min}.
The slope of rolling potential and effective backreaction potential is
\bea
\frac{V'_{\rm roll}}{\Lbr^4}  &=& -1 \,,
\\
\frac{\widetilde V'_{\rm br}}{\Lbr^4} &=&  \frac{v_m^2(\theta)}{\vew^2} 
\left(\sin\theta - \frac{\vew^2}{\Lambda^2} \right)\,,
\eea
where the prime denotes a derivative with respect to $\theta$. 
Consider an integer number $m$ such that $v_m^2 \lesssim \vew^2$. 
We see that the slope of the potential $V'_{\rm eff}(\theta) <0$ for $\theta \in [ 0 ,2\pi)$. The rolling potential provides a constant driving force, and eventually, the integer number changes by one unit $m \to m+1$. 
This unit change of the integer number $\Delta m = +1$ results in an incremental change of the backreaction potential as $\Delta V'_{\rm br} / \Lbr^4 = \Delta v^2 /\vew^2 = \pi  \mu_b^2 / \Lambda^2 = \pi \delta^2 $. 
This sequence continues to happen until one finds an integer number such that $v_m\simeq \vew$, allowing for a solution of $V'_{\rm eff} = 0$ in $\theta \in [0,2\pi)$ for the first time.

To investigate the relaxion potential near local minima, let us define $m_\star$ as the smallest integer number that allows solutions to $V'_{\rm eff} (\theta) = 0$. 
Unless there is an accident, there are generally two solutions for $\theta$, one corresponds to the local minimum, and the other corresponds to the local maximum. 
To see where the relaxion stops, it is important to note 
\bea
\max_{m = m_\star} V_{\rm eff}'(\theta) / \Lbr^4 \sim \delta^2\,,
\label{maxVp}
\eea
where the left hand side represents the maximum value of $V_{\rm eff}'(\theta)/\Lbr^4$ for $\theta \in [ 0, 2\pi)$ with a given $m=m_\star$.
This is because the change of $V_{\rm eff}'(\theta)$ over $\Delta\theta = 2\pi$ is given by $\Delta V_{\rm br}'(\theta)/\Lambda_{\rm br}^4 = \pi \delta^2$, so for the first time $V_{\rm eff}'(\theta)$ can exceed zero, it is at most by that amount.
Note that the maximum of $V'_{\rm eff}(\theta)$ occurs when $V_{\rm eff}''(\theta_\star) = 0 $,
\bea
0 = V_{\rm eff}''(\theta_\star) \simeq \Lbr^4 \Big[ \cos\theta + {\cal O}(\eps) \Big]\,,
\label{inflection}
\eea
leading to an estimation $\theta_\star \sim \pi/2 + {\cal O}(\eps)$.
The $\eps$-correction appears due to the relaxion-dependence of the Higgs vacuum value, i.e. $\eps \simeq \partial \ln v^2 /\partial \theta$.
The above two equations Eqs.~\eqref{maxVp}--\eqref{inflection} provide important information on the local extremum of relaxion potential: the local minimum and maximum are separated by $\Delta \theta \sim \delta$, and they are centered around $\theta_\star = \pi/2 + {\cal O}(\eps)$ at which the second derivative of the potential vanishes. 
Being $\delta$-close to this inflection point, the relaxion mass is naturally suppressed by $\delta$.
A schematic figure of $V'_{\rm eff}(\theta)$ near $\theta_\star$ can be found in Fig.~\ref{fig:sketch}.

A more careful investigation on the local shape of relaxion potential can be done by expanding the potential around $\theta_\star$.
We find an approximate form of $V'_{\rm eff}(\theta)$ around $\theta_\star$ as
\bea
V_{\rm eff}'(\theta) = V_{\rm eff}'(\theta_\star) + \frac{1}{2} V_{\rm eff}'''(\theta_\star) ( \theta-\theta_\star)^2 + \cdots,
\eea
where each coefficient is given as
\bea
\frac{V_{\rm eff}'(\theta_\star)}{\Lbr^4} &\sim& {\cal O}(\delta^2)\, ,
\\
\frac{V_{\rm eff}'''(\theta_\star)}{\Lbr^4} &\sim& {\cal O}(1)\, .
\eea
The first local minimum and maximum is separated by
\bea
\Delta \theta =2^{3/2} \sqrt{\frac{V_{\rm eff}'}{V_{\rm eff}'''}} \sim 2 \delta\, ,
\eea
while the mass at the first local minimum is
\bea\label{eq:relaxedmass}
m_\phi^2 = V'''_{\rm eff}(\theta_\star) (\theta_{\rm min} - \theta_\star) \simeq \delta\,  \frac{\Lbr^4}{f^2}\, .
\eea
We see that the squared mass is indeed suppressed by additional small parameter $\delta = \mbr/\Lambda$ compared to the naive expectation $\Lbr^4/f^2$.
If one approximates the relaxion potential around the local minimum as a quadratic potential, one can immediately find that the potential height is also suppressed by $\delta^3$ relative to $\Lbr^4$, 
\bea
\Delta V_{\rm eff} 
= \int_{\theta_\star - \frac{\Delta\theta}{2}}^{\theta_\star + \frac{\Delta\theta}{2}} d\theta \,\, V'_{\rm eff}(\theta)
\sim \delta^3  \Lbr^4 \, . 
\label{potential_height}
\eea 
One can repeat the similar analysis for subsequent minima with $m>m_\star$, and all results hold upon substituting $\delta \to \sqrt{n} \delta$, with $n = m-m_\star +1$ representing $n$-th local minimum in relaxion field space.
This result can be easily obtained by noting that $\max_n V'(\theta)/\Lbr^4 \sim n \delta^2$. 

The backreaction potential under consideration should be understood as a leading order term; there might exist subleading higher harmonics.
Such higher harmonics do not change the qualitative picture, because the crucial point for the parametric suppression of mass is that the Higgs VEV changes incrementally, $\Delta v^2 /v^2 = \pi \delta^2$.
The squared relaxion mass is naturally suppressed by $\delta$, no matter what periodic backreaction potential we impose for the relaxion scenario, as long as the change of the backreaction potential is controlled by the same $\delta$.

\subsection{Classical evolution}\label{sec:classical}
The underlying assumption in the previous section is that the Higgs adiabatically follows its instantaneous minimum, and that the relaxion is slow-rolling such that it stops at $V'_{\rm eff}(\theta) = 0$. 
We clarify the conditions under which our assumption can be justified.

\begin{figure}
\centering
\includegraphics[scale=0.35]{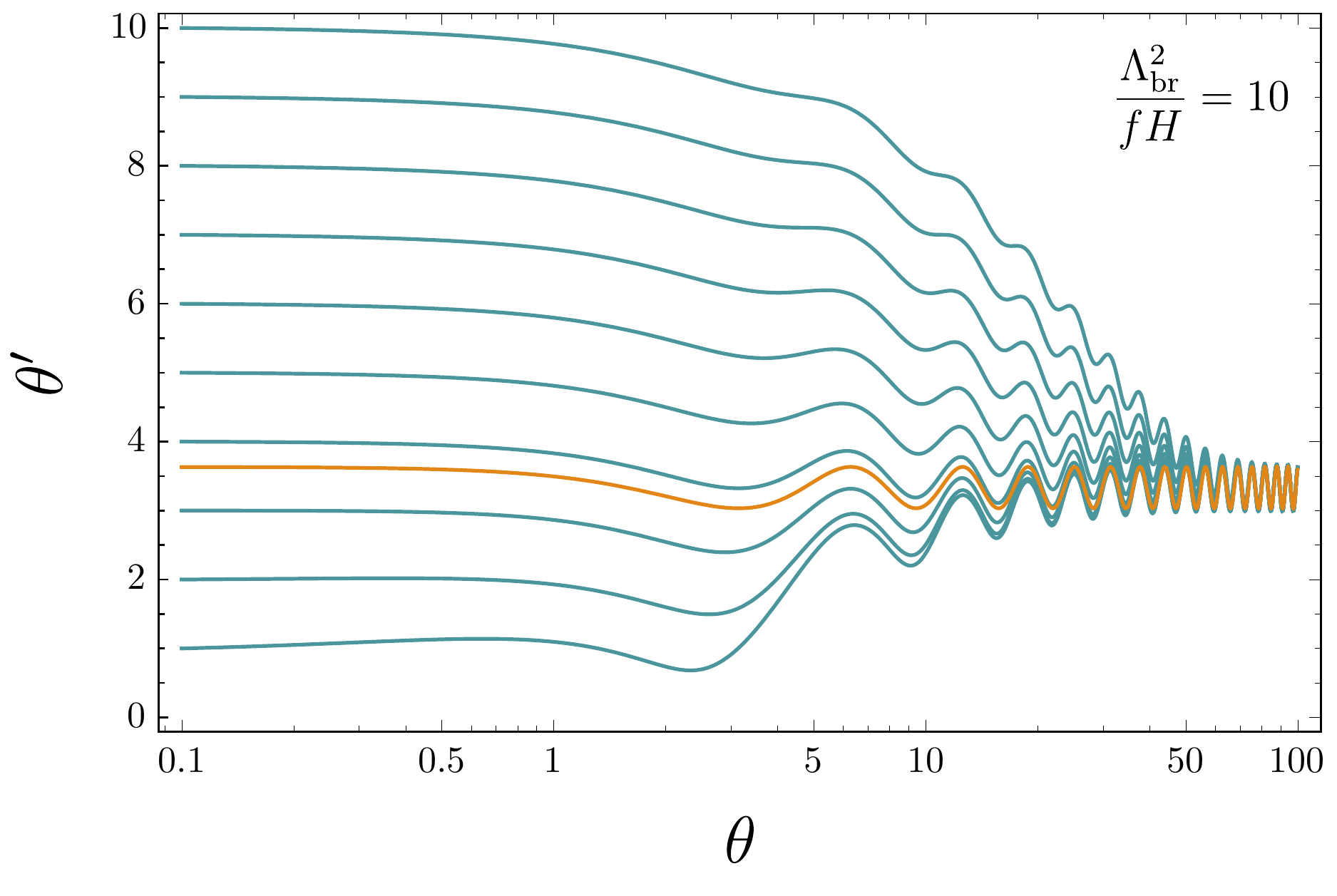}
\quad
\includegraphics[scale=0.35]{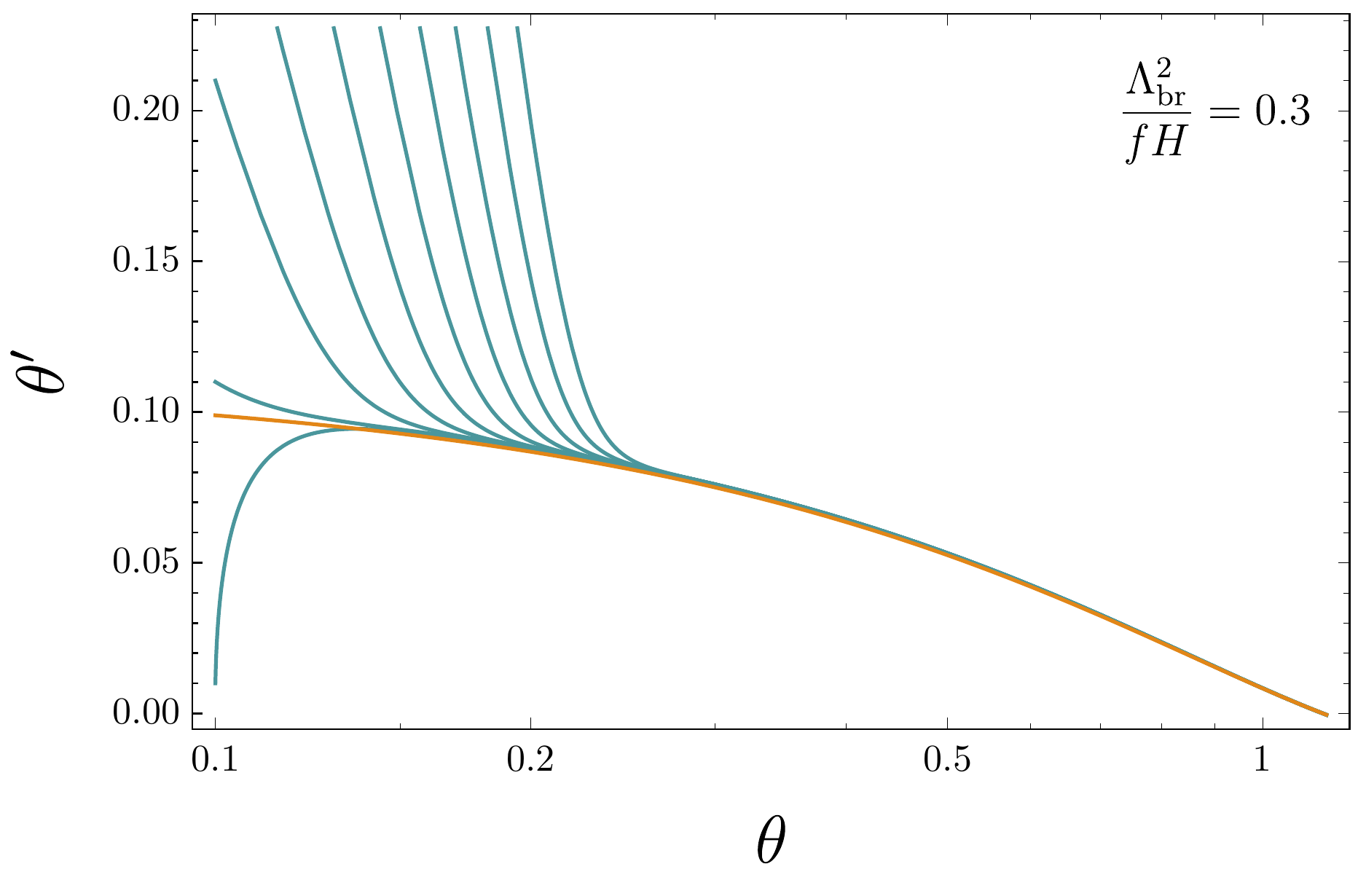}
\caption{Attractor behavior of the solution in phase diagram. 
The prime denotes a derivative with respect to dimensionless time parameter $(\Lbr^2/f) t$. 
(Left) The phase diagram when the curvature is larger than the Hubble, $V''/H^2 >1$.
(Right) The phase diagram when the curvature is smaller than the Hubble, $V''/H^2<1$.
It is clear that the attractor behaviour for $V'' / H^2 >1$ shows unbounded evolution in the phase diagram.
The blue lines correspond to the solution with different initial conditions.
The orange lines in both figures show attractor solutions. 
For $V'' / H^2>1$, it corresponds to Eq.~\eqref{attractor1}, while, for $V'' /H^2 <1$, it corresponds to the slow-roll approximation, $\dot{\phi} = - V' /3H$.  }
\label{fig:attractor}
\end{figure}

The discussion has to do with the relative size of the curvature of the relaxion potential and inflationary Hubble parameter.
When the backreaction is turned on, the slow-roll approximation is valid only when the curvature of the potential is smaller than the squared Hubble expansion parameter. 
One can see this by substituting the slow-roll approximation, $\dot{\phi} = - V' / 3 H$, to the equation of motion, $\ddot{\phi}+ 3 H \dot{\phi} + V'=0$, and observe that the relative correction is ${\cal O}(V''/H^2)$.
This shows that the slow-roll approximation is appropriate only for $V'' / H^2  \sim \Lbr^4/ f^2 H^2 < 1$.
Alternatively, this condition can be written as $ t \simeq f / \Lbr^2 > 1/H$, where $t$ is the time scale that the relaxion evolves for $\Delta \phi = 2 \pi f$. 
For $V'' / H^2 >1$, this time scale $t \sim f / \Lbr^2$ becomes smaller than the Hubble time scale, $t_H \sim H^{-1}$. 
This indicates that the Hubble friction is insufficient to dissipate the kinetic energy obtained due to the potential difference of backreaction potential, so that the relaxion evolves as if there is no wiggly backreaction potential.
In other words, the wiggly backreaction potential is effectively averaged out, and does not provide required backreaction to the relaxion evolution. 
Specifically, we find an attractor solution for $\Lbr^2 / f H >1$ as
\bea
\theta' = \frac{\Lbr^2 }{3Hf } + \frac{3Hf}{\Lbr^2} \cos\theta + {\cal O}\left( \frac{H^2 f^2}{\Lbr^4} \right),
\label{attractor1}
\eea
where the prime denotes a derivative with respect to dimensionless time parameter, $(\Lbr^2 /f)t$. 
According to the attractor solution, the relaxion shows unbounded evolution in the phase diagram, signaling that it overshoots the minimum due to inertia obtained from the relaxion potential.
In Fig.~\ref{fig:attractor}, one can clearly observe this attractor behavior regardless of initial conditions. 
Due to this reason, we focus on $\Lbr^2 / f H < 1$ in what follows.\footnote{This condition is briefly discussed in~\cite{Choi:2016luu}. 
Although we are interested in the original relaxation scenario described in~\cite{Graham:2015cka}, it is worth mentioning that a recent study~\cite{Fonseca:2019ypl,Fonseca:2019lmc} has shown that for relaxion mass $\mathcal{O}(\gtrsim \keV)$, the relaxation of electroweak scale could also happen for $\Lbr^2 /f > H$ through a larger barrier potential or through relaxion particle production. It is discussed that, in such cases, the relaxion mass and mixing angle would follow naive expectation Eqs.~\eqref{eq:mphinaive1}--\eqref{eq:naivemixing}.} In other words, we are interested in $m_\phi < \Lbr^2 / f < H$.

\subsection{Quantum evolution}
The relaxion is classically stabilized at the shallow part of the potential. 
Even though the local minimum is classically stable, the relaxion could further evolve  through quantum mechanical processes in the inflationary universe. 
Since the present value of relaxion mass depends on the minimum where the relaxion is stabilized at the end of inflation, we investigate in more detail the final relaxion minimum, including the quantum evolution during inflation.
For the relaxion mass smaller than the Hubble parameter, most relevant processes would be transitions due to Hawking-Moss instanton~\cite{Hawking:1981fz}, or stochastic quantum dispersion, which leads to identical conclusion~\cite{Starobinsky:1986fx,Linde:1991sk}. 
Note that Coleman-de Luccia (CdL) transition~\cite{Coleman:1980aw} does not exist for the case under consideration, $m_\phi < \Lbr^2/f <  H_I$~\cite{Hawking:1982my}.

The Hawking-Moss instanton describes thermal hopping of the relaxion field from the local minimum to the local maximum, and hence, allowing the relaxion to access to the next local minimum.
The rate at which this thermal hopping happens can be characterized by the tunnelling rate (per volume) $\Gamma = A\,e^{-B}$, where $A$ is a mass-dimension four coefficient and $B = S_E[\phi_{\rm max}] - S_E[\phi_{\rm min}]$ is the difference of Euclidean action at the local maximum of the potential $S_E[\phi_{\rm max}]$ and at the local minimum $S_E[\phi_{\rm min}]$.
We will not be interested in the coefficient $A$, but focus on the exponent $B$.

The Hawking-Moss instanton action can be trivially computed since the inflation is driven by a separate inflaton sector, indicating that the Hubble parameter barely depends on the energy density of relaxion sector. 
We find the exponent $B$ as
\bea
B = \int d^4x \sqrt{g} \left[V(\phi_{\rm max}) - V(\phi_{\rm min}) \right]
= \frac{8\pi^2}{3} \frac{ \Delta V}{H_I^4}\,,
\label{HM}
\eea
where $8\pi^2/3 H_I^{4}$ is the surface area of four-sphere with radius $H_I^{-1}$. 
Since the tunnelling would happen until the tunnelling probability is exponentially suppressed, which starts when $B \sim \mathcal{O}(1)$, we can estimate at which minimum the relaxion is finally stabilized as
\bea
n = \max\left[1, \, \left( \frac{3}{8\pi^2 \delta^3} \frac{H_I^4}{\Lbr^4}\right)^{\frac{2}{3}} \right].
\label{final_n}
\eea
where we have used Eq.~\eqref{potential_height} for the potential barrier $\Delta V$ at $n$-th minimum, $\Delta V = (\sqrt{n} \delta)^3 \Lbr^4$.  
Note that if $\Delta V > H_I^4$ at the first local minimum, the relaxion would be stuck at the first minimum until present.

The same result can be obtained by considering the stochastic evolution of the relaxion. 
Since the relaxion mass is smaller than the Hubble expansion parameter as we discussed in the previous section, the relaxion experiences stochastic evolution $\Delta \phi \sim H_I / 2\pi$ for a unit e-folding number. 
The probability distribution functional $P[\phi,t]$ for the relaxion field evolves according to the Fokker-Planck equation~\cite{Starobinsky:1986fx, Rey:1986zk, Linde:1991sk},
\bea
\frac{\partial P}{\partial t} = \frac{1}{3H_I} \frac{\partial}{\partial \phi} \left( \frac{\partial V}{\partial \phi} P \right) + \frac{H_I^3}{8\pi^2} \frac{\partial^2 P}{\partial \phi^2}\,.
\eea
If one approximate the potential around local minima as $V = \frac{1}{2} m_\phi^2 \phi^2$, the solution can be easily obtained as 
\bea
P[\phi,t] = \frac{1}{\sqrt{2\pi \sigma^2(t)} } \exp\left[ - \frac{[\phi - \mu_\phi (t)]^2}{2\sigma_\phi^2(t)} \right].
\eea
Here, $\mu_\phi(t) = \langle \phi(t) \rangle$ and $\sigma_\phi^2(t) = \langle \phi^2(t) \rangle -\langle \phi(t) \rangle^2 $ are mean and variance of the relaxion field $\phi$ coarse-grained over a Hubble size $\sim 1/H_I$,
\bea
\mu_\phi(t) &=& \mu_\phi(t_0) \exp\left[ - \frac{m_\phi^2}{3H_I^2} N \right],
\\
\sigma_\phi^2(t) &=& \frac{3H_I^4}{8\pi^2 m_\phi^2} \left[ 1 - \exp\left( - \frac{2m_\phi^2}{3H_I^2} N \right) \right],
\eea
where $N = H(t-t_0)$ is the number of e-folding.

Note that the required number of e-folding to scan the electroweak scale is $N  \sim (H_I f / \Lbr^2)^2 ( \Lambda/\Lambda_{\rm br})^4$.
If one assumes that there is similar number of e-folding after the relaxion is stabilized around one of its local minima, then the mean and variance of coarse-grained field can be approximated as $\mu_\phi \simeq 0$ and $\sigma_\phi^2 \simeq 3H_I^4 / 8\pi^2 m_\phi^2$. 
The probability density functional becomes
\bea
P[\phi,t] \propto \exp \left[ - \frac{8\pi^2 V}{3H_I^4}  \right].
\eea
The exponential dependence coincides with that of Hawking-Moss instanton~\eqref{HM}. 
Given that the distance between local minimum and maximum at $n$-th minimum is $\Delta \theta \sim \sqrt{n} \delta$, we arrive at the same conclusion~\eqref{final_n}.
This is the minimum where the relaxion is stabilized until the present universe.

\section{Naturalness}\label{sec:low_obs}
While the relaxion mass is parametrically suppressed, the Higgs-relaxion mixing angle remains unmodified compared to its naive value. 
Consequently, the measured value  
of the mixing angle can be puzzling for low energy observers, in the context of naturalness, as follows. Suppose that, in one of the experiments discussed in Section~\ref{sec:pheno}--\ref{sec:pheno_low}, we were able to observe a signal of a new light scalar $\phi$ with a mass $m_\phi^2 \ll m_h^2$ and measure its couplings to the SM states. We can interpret this signal in terms of the minimal relaxion framework, a la {\it Higgs portal}. 
For a small mixing angle, the leading interactions are of the following form (we omit all numerical order-one factors in this section and use $m_h$ and $\vew$ interchangeably)
\bea\label{eq:hipo}
\sin \theta_{h \phi}\, \vew |H|^2 \phi\,.
\eea    
The corresponding mass matrix of the relaxion and the Higgs boson $h$ reads
\bea\label{eq:mmbottomup}
{\cal M}^2 = \left[
\begin{array}{cc}
m_h^2 &  m_h^2 \sin \theta_{h \phi}
\\[6pt]
m_h^2 \sin \theta_{h \phi}  & m_{\phi0}^2
\end{array}
\right],
\eea
where the bare $\phi$ mass, $m_{\phi0}^2$, is reconstructed from the measured $m_\phi^2$ and $\sin \theta_{h\phi}$ using the relation $\det {\cal M}^2 = m_h^2 m_\phi^2$, which gives
\bea\label{eq:mhtun}
m_{\phi}^2 = m_{\phi0}^2 -  m_h^2 \sin^2 \theta_{h \phi}\,.
\eea
The left hand side ({\it lhs}) positivity requires that $m_{\phi0}^2 $ is greater than $m_h^2 \sin^2 \theta_{h \phi}$. Furthermore, the natural expectation is that there is no fine cancellation between the two terms. We would then find $m_{\phi}^2 \simeq m_{\phi0}^2$, and arrive at a prediction which is generic for the Higgs portal models~\cite{Piazza:2010ye, Arvanitaki:2015iga, Graham:2015ifn},
\bea\label{eq:hiportal}
\sin \theta_{h \phi} \lesssim \frac{m_{\phi}}{m_h}\, .
\eea

However, we already know that the relaxion mass is suppressed with respect to the naive estimate. We therefore  expect to observe a significant cancellation between the two tree-level contributions to the relaxion mass in the right hand side ({\it rhs}) of Eq.~(\ref{eq:mhtun}). To see how this happens, we compute the entries of the mass matrix directly, using the results of the previous section,
\bea
m_{\phi0}^2 &=& \partial^2_\phi V(h,\phi) = \frac{\Lbr^4}{f^2} \cos \theta_0 \simeq \frac{\Lbr^4}{f^2} \left( \eps + \sqrt n \delta \right), 
\label{mphi0}
\\
m_h^2 \sin \theta_{h \phi}  &=& \partial_\phi \partial_h V(h,\phi) \simeq \frac{\Lbr^4}{\vew f} \sin \theta_0 \simeq \frac{ \Lbr^4}{\vew f}\,.
\eea
Since two small parameters that we have defined in the introduction appear frequently, we remind readers their definition again
$$
\delta \equiv \frac{\mbr}{\Lambda}\,,
\qquad
\eps \equiv \frac{\mbr^2}{\vew^2}\,.
$$
With the derived expression for the physical relaxion mass
\be
m_\phi^2 \simeq \sqrt n \delta \frac{\Lbr^4}{f^2}\, ,
\label{mphi}
\ee
we find the ratios between the contributions to $m_\phi^2$, and $m_\phi^2$ itself
\bea
\frac{m_{\phi0}^2}{m_\phi^2}  &\sim& 
1+ \frac{\eps}{\sqrt{n}\delta}\,,
\\
\frac{ m_h^2 \sin^2 \theta_{h \phi}}{m_\phi^2} &\sim&
\frac{\eps}{\sqrt{n}\delta}\,,
\eea
which signal a presence of fine tuning if 
\bea\label{eq:tuning}
\frac{\eps}{\sqrt{n}\delta} = \frac 1 {\sqrt n} \frac{\Lambda \mbr}{\vew^2} >1\,.
\eea
Furthermore, we find that the relation between the mixing angle and the mass 
\bea
\sin^2 \theta_{h \phi} = \frac{\eps }{\sqrt{n}\delta } \frac{m_\phi^2}{m_h^2} \label{sintune}
\eea
violates the naturalness bound~(\ref{eq:hiportal}) if the inequality~(\ref{eq:tuning}) holds.

It is interesting to notice that, despite the fact that the relaxion mass in the first minimum is always suppressed by a small factor $\mbr/\Lambda$ with respect to the naive estimate $(m_\phi^2)_{\rm naive}=\Lbr^4/f^2$, an additional condition (\ref{eq:tuning}) is still needed for $m_\phi^2$ to appear tuned. This happens because the actual bare mass $m_{\phi 0}^2 \simeq (\Lbr^4 /f^2) \cos \theta_0$, with respect to which the tuning is measured, carries a suppression from  $\cos \theta_0 \sim \eps$ if the relaxion is close to $\pi/2$.

As a next step, we may try to analyse whether the tuned values of the tree-level parameters are stable under radiative corrections. The largest such a correction is a contribution of the Higgs portal interaction~(\ref{eq:hipo}) to the bare $\phi$ mass
\bea
\Delta m_{\phi 0}^2 \sim \frac{\sin^2 \theta_{h \phi}}{16 \pi^2}\, \vew^2\,,
\eea
which has the same parametric form as the $m_h^2 \sin^2 \theta_{h \phi} $ contribution to the relaxion mass in Eq.~(\ref{eq:mhtun}) and implies the same tuning, up to a loop factor.

In summary, the combination of the relaxion mass and couplings, once observed, may seem hard to be reconciled with naturalness without knowing the global structure of the scalar potential. By solving the naturalness problem of the SM with the relaxion, we may end up with two unnaturally-looking scalars -- the Higgs and the relaxion.

\section{Mixing angle and phenomenology}\label{sec:mixing_angle}
In this section, we discuss relaxion phenomenology paying particular attention to the implications of the relaxion mass suppression. The mass matrix of the Higgs and relaxion is
\be
{\cal M}^2=\left[\begin{matrix}
m_h^2 & \sqrt 2 \frac {\Lbr^4}{\vew f} \sin \theta_0 \\[8pt]
\sqrt 2 \frac {\Lbr^4}{\vew f} \sin \theta_0 & \frac {\Lbr^4}{f^2} \cos \theta_0
\end{matrix}\right].
\ee
Using the fact that the Higgs mass is always greater than the off-diagonal entries of ${\cal M}^2$ and the relaxion mass, we associate $({\cal M}^2)_{11}$ with the physical Higgs mass, while the relaxion-Higgs mixing angle is approximately given as
\bea\label{mixing_angle}
\sin \theta_{h\phi} \simeq \sqrt{2} \frac{\Lbr^4}{f \vew m_h^2} \sin\theta_0 \simeq \sqrt{2} \frac{\Lbr^4}{f \vew m_h^2}\,.
\eea
Since $\theta_0 \simeq \pi/2$ for the most part of the parameter space, we take $\sin\theta_0 = 1$ in the following.
This mixing angle is identical to Eq.~\eqref{sintune}, and this can be easily shown by substituting the relaxion mass Eq.~\eqref{mphi} to Eq.~\eqref{sintune} and using $\eps = (\mu_b/\vew)^2$.
We also use $\Lbr^2 = \mbr \vew$ instead of $\mbr$ when possible.
 
The size of the mixing angle is the main parameter, relevant for low energy phenomenology and the experimental searches. The size depends on the allowed range of parameters $\Lambda, f$ and $\Lbr$. The range of their values is restricted as
\bea
\Lambda > \Lambda_{\rm min} = 1\TeV\,,
\qquad
f \geq \Lambda \,,
\qquad
\Lbr \leq \vew\,,
\label{bound_params}
\eea
and, in most minimal cases, one can also impose $f < \Mp$. 
These restrictions are correlated in a non-trivial way through the relaxion mass
\bea\label{eq:masspheno}
m_\phi^2 
= \sqrt{n}\, \delta \,\frac{\Lbr^4}{f^2}
= \sqrt{n}\, \frac{\Lbr^2}{\Lambda\vew} \frac{\Lbr^4}{f^2}\,.
\eea
Since we are interested in finding the maximum and minimum value of the mixing angle for a fixed relaxion mass, we express the mixing angle in terms of relaxion mass as
\bea
\sin\theta_{h\phi} = \left( \frac{m_\phi^4 f \Lambda^2}{n \vew^7} \right)^{1/3}\!\!.
\label{ma_re}
\eea
Therefore, we need to find the value of $f$ and $\Lambda$ which maximizes and minimizes the mixing angle, while satisfying the constraints Eq.~\eqref{bound_params}.  
A successful cosmological relaxation of the Higgs mass implies yet another constraint on the model parameters~\cite{Graham:2015cka}
\bea
\frac{\Lambda^2}{\Mp} < \left( \frac{\Lbr^4}{f} \right)^{1/3}\!\!,
\label{eq:relaxation_condition}
\eea
which arises from combining two conditions, $H_I > \Lambda^2 / \Mp$ (inflaton sector dominates the total energy density), and $H_I < (\Lbr^4/f)^{1/3}$ (classical evolution dominates quantum evolution). 
The condition $H_I> \Lbr^2/f$ discussed in Sec.~\ref{sec:classical} can be trivially satisfied. 
We show below that the above constraints lead to both upper and lower bound of the relaxion-Higgs mixing angle.

\subsection{Maximum mixing angle}\label{sec:maxmix}

To find the maximum mixing angle, we must find the maximum value of $f \Lambda^2$ for a fixed relaxion mass. 
We ignore any order one numerical coefficients for the analytic estimations below, while we use precise estimates for the plots presented in the following. 
Using the relaxion mass expression Eq.~\eqref{eq:masspheno}, one finds $ f^2 = \sqrt{n} \Lbr^6/ (\Lambda \vew m_\phi^2)$, and by substituting this expression to Eq.~\eqref{ma_re}, we see that the maximum mixing angle is realized for the maximum $\Lambda$ and $\Lbr$. 
From the constraints Eq.~\eqref{bound_params}, we also find $\Lambda^2 < f^2 < \sqrt{n} \vew^5/ (\Lambda  m_\phi^2)$.
Setting $\Lambda$ to its maximum value $\Lambda = (\sqrt{n} \vew^5 /m_\phi^2)^{1/3}$, while choosing corresponding $f$, we find the maximum mixing angle as
\bea\label{eq:mixmax1}
 \sin \theta_{h\phi} < 
 \frac{1}{n^{1/6}} \left( \frac{m_\phi}{\vew}\right)^{2/3}\! , 
\eea
which provides the maximum mixing angle for the relaxion mass above eV. 

The above limit is obtained by using Eqs.~\eqref{bound_params}-\eqref{eq:masspheno}. 
As the relaxion mass decreases, the cosmological constraint Eq.~\eqref{eq:relaxation_condition}, places a more stringent limit on the allowed range of $f$ and $\Lambda$.
Combining Eqs.~\eqref{bound_params}--\eqref{eq:relaxation_condition}, we find $ n \Lambda^{16}/ (m_\phi^4 \Mp^9 \vew^2) < f < [\sqrt{n}\vew^5 / (\Lambda m_\phi^2)]^{1/2}$, where the lower bound now arises from cosmological consideration~\eqref{eq:relaxation_condition}.
Using these inequalities, we choose the maximum value of $\Lambda$ and corresponding $f$, and find 
\bea\label{eq:mixmax2}
\sin \theta_{h\phi} 
< \left( \frac{m_\phi^{4} \Mp}{n \vew^{5}} \right)^{3/11}\!,
\eea
which yields the maximum mixing angle for the relaxion mass below eV.

For even smaller mass, the relaxion decay constant eventually becomes super-Planckian. One may also limit the decay constant to be sub-Planckian, $n \Lambda^{16}/ (m_\phi^4 \Mp^9 \vew^2) <f < \Mp$, leading to the upper bound on $\Lambda$. 
This requirement limits the mixing angle as
\bea\label{eq:mixmax3}
\sin \theta_{h\phi} 
< \left(\frac{m_\phi^{2} \Mp}{\sqrt{n} \vew^{3}} \right)^{3/4}\!\!,
\eea
which is dominant for relaxion mass below $10^{-8}\eV$. These above bounds [Eqs.~\eqref{eq:mixmax1}--\eqref{eq:mixmax3}] are graphically represented in Fig.~\ref{fig:min_max_numerical}

\subsection{Minimum mixing angle}\label{sec:minmix}

The lower bounds on the mixing angle are of two types, relaxion mass dependent, and independent ones.   
Using Eq.~\eqref{ma_re} and $f\geq\Lambda>\Lambda_{\rm min}$, we find 
\bea\label{eq:mixmin1}
\sin \theta_{h\phi} 
> \left( \frac{m_\phi^{4}\Lambda_{\rm min}^3}{n \vew^{7}} \right)^{1/3}\! ,
\eea
which depends on relaxion mass. Similarly to the maximum mixing angle, there is also a minimum mixing angle obtained from the relaxation condition Eq.~\eqref{eq:relaxation_condition}. 
This leads to $\sin\theta_{h\phi} > \left( \frac{\Lambda_{\rm min}^2 }{ \vew \Mp} \right)^3 \simeq 10^{-44}$. 
Note that this lower bound is independent of $n$ and $m_{\phi}$. These bounds have been graphically represented in Fig.~\ref{fig:min_max_numerical}.

\subsection{Comparison with the naive generic case }\label{sec:genmix}

It is interesting to compare the derived relaxion mixing angles with those in the  case where the physical relaxion mass is not suppressed with respect to the naive estimate $\Lbr^4 /f^2$. This value of mass corresponds to a local minimum where $\sin \theta_0 \sim \cos \theta_0 \sim {\cal O}(1)$, or in terms of relaxion terminology, it corresponds to $n \sim 1/ \delta^2$, which is unlikely to be realized given Eq.~\eqref{final_n} (with alternative stopping mechanism, the relaxion may stop at this minimum  for $m_{\phi}\gtrsim \mathcal{O}(\keV)$~\cite{Fonseca:2019lmc}). 

\begin{figure}
\centering
\includegraphics[scale=0.3]{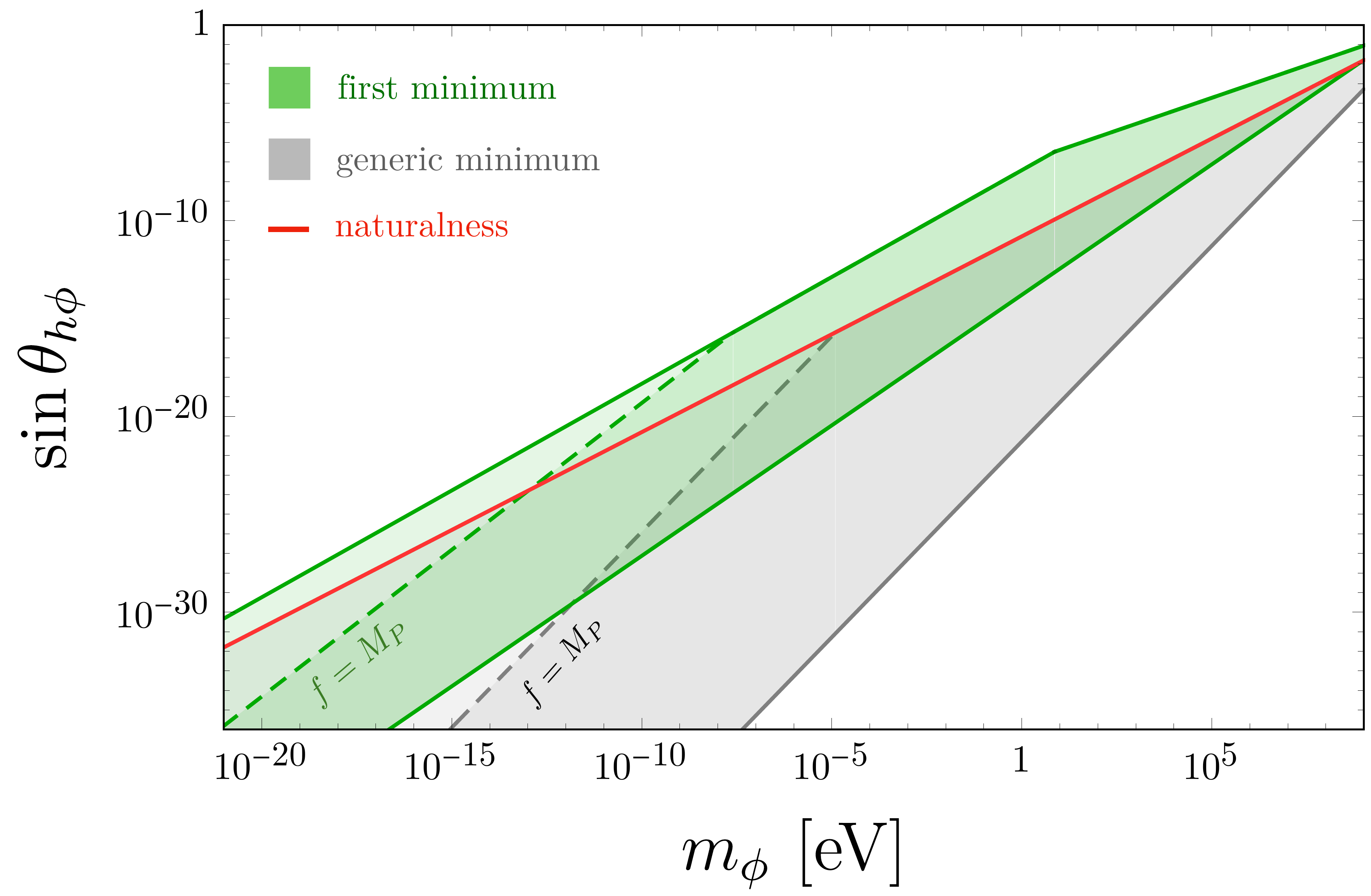}
\caption{Relaxion-Higgs mixing angle as a function of the relaxion mass. The green band corresponds to the first minimum, the gray band to the generic minimum, and the red line is the naturalness bound of Eq.~\eqref{eq:hiportal}. The regions above the dashed line correspond to super-Planckian relaxion decay constant.}
\label{fig:min_max_numerical}
\end{figure}

As we have already discussed in the introduction, in the naive case, the relaxion mass in this minimum is given as  
\be\label{eq:relaxmassgen}
m_\phi^2 \simeq \frac{\det {\cal M}^2}{m_h^2}  
\sim \frac{\Lbr^4}{f^2}\, ,
\ee 
while the mixing angle is 
\be\label{eq:sinmixgen}
\sin \theta_{h \phi} \simeq \frac {\Lbr^4}{f \vew^3}\, .
\ee
The mixing angle can be written in terms of unsuppressed relaxion mass as

\be
\sin \theta_{h \phi} \simeq \frac{m_\phi}{\vew} \frac{\Lbr^2}{\vew^2}\,.
\ee 
Since $\Lbr < \vew$ for the relaxation scenario~\eqref{bound_params}, we find the maximum mixing angle in this generic relaxion minimum as 
\be 
 (\sin \theta_{h \phi})_{\rm naive} 
 < \frac{m_\phi}{\vew}\, .
\ee
This coincides with the naturalness bound, which was discussed in Section~\ref{sec:low_obs}. 
By expressing the mixing angle in terms of $f$,
\be
\sin \theta_{h \phi} \simeq \frac{m_\phi^2 f}{\vew^3}\, ,
\ee
we derive an upper and a lower bound on $\sin \theta_{h \phi}$ for this case,
\be
\frac{m_\phi^2 \Lambda_{\text{min}}}{\vew^3} < (\sin \theta_{h \phi})_{\rm naive}< \frac{m_\phi^2 \Mp}{\vew^3}\, .
\ee
Notice that, if $f < \Mp$ constraint is imposed, the relaxion-Higgs mixing in the generic case \emph{cannot} reach the naturalness line $\sin \theta_{h \phi} = {m_\phi}/{\vew}$ if $m_\phi < \vew^2/\Mpl$. In Fig.~\ref{fig:min_max_numerical}, we show the analytic results derived in Sec.~\ref{sec:maxmix}--\ref{sec:genmix}. 

\subsection{$h-\phi$ interactions}\label{sec:xquartic}  

It is of a phenomenological interest to derive the couplings leading for the Higgs-relaxion interactions, of which the most important one is $h \phi^2$. It can contribute to the Higgs decays to the relaxion, as well as to the relaxion production via off-shell Higgs.    
Before rotation to the mass eigenstates, the relevant terms of the Lagrangian are 
\bea
{\cal L}_{h \phi^n} &=& - {\lambda} |H|^4 + \mbr^2 |H|^2 \cos \phi/f \label{eq:hphin}\\ 
&\underset{h\to\phi^2}{\to}& 
-\sqrt 2 \lambda \vew \, h^3  
- \frac{\Lbr^4}{2\vew^2 f}\sin \theta_{0} \,h^2 \phi 
- \frac{\Lbr^4}{\sqrt 2 \vew f^2}\cos \theta_{0} \,h \phi^2  \label{eq:hphi2}
\eea
where we have expanded $\phi$ around its vacuum expectation value, $\phi/f \to  \theta_0 + \phi/f$, and only kept the terms $h^m \phi^n$ with $m+n=3$, which give rise to $h \phi^2$ couplings after mass diagonalization. We have omitted the coupling $g \Lambda |H|^2 \phi$ from the Lagrangian \eqref{eq:hphin}, as its resulting contribution to $h \phi^2$ is always suppressed by $\vew^2/\Lambda^2$ with respect to the rest. We also do not include the operator $\phi^3$ to the expansions~(\ref{eq:hphi2}), for its effect being additionally suppressed by $\vew/f$.

Before moving further, it is interesting to comment on a connection to the Higgs portal models. In section~\ref{sec:low_obs}, we have already pointed out that the relaxion-Higgs mixing angle violates the natural Higgs portal prediction. This mixing can be related to the cubic interaction term $\phi |H|^2$ in the general Higgs portal parametrization. Another term present in these models is the cross-quartic $\phi^2 |H|^2$, which, together with the mixing angle, determines the size of the $h\phi^2$ coupling in the mass eigenstate basis. As the cross-quartic contributes to the $\phi$ mass, its coefficient has to satisfy the naturalness criterion 
\be\label{eq:quartnat}
\lambda_{H \phi} \lesssim m_\phi^2/\vew^2\,. 
\ee
In the relaxion case we have, instead, 
\be
\lambda_{H \phi} \sim \frac{\Lbr^4}{\vew^2 f^2}\cos \theta_{0} \sim {m_{\phi0}^2}/{\vew^2}  \sim 
\left(1+ \frac{\eps}{\sqrt{n}\delta}\right) m_\phi^2/\vew^2\sim m_\phi^2/\vew^2+ \sin^2 \theta_{h \phi}\,, 
\ee
where as can be seen directly from Eq.~\eqref{sintune}, it therefore violates the naturalness bound of Eq.~\eqref{eq:quartnat} in the dynamical tuning regime~\eqref{eq:tuning}.

To find Higgs-relaxion interactions in the mass eigenbasis, we perform a rotation 
\bea
\begin{cases}
\;\phi  \to  \cos \theta_{h \phi}\;  \phi + \sin \theta_{h \phi}\;  h \, , \\
\;h  \to  \cos \theta_{h \phi}\;  h  - \sin \theta_{h \phi}\;  \phi \, .
\end{cases}
\eea 
For the $h\phi^2$ coupling, we obtain
\bea
{\cal L}_{h \phi^2} 
&\simeq&
- \frac{\Lbr^4}{\sqrt{2} \vew f^2}
\left( 
\cos \theta_0 +
\frac{\eps}{4} \sin^2 \theta_0 \right) \, h \phi^2  \label{eq:hphi2func} \\
&\simeq&
- \frac{\Lbr^4}{\sqrt{2} \vew f^2}
\left(
\sqrt{2\pi n} \, \delta
+ \frac{3\eps}{4} 
\right) \, h \phi^2\,. \label{eq:hphi2param}
\eea
where we have used $\sin\theta_0 \simeq 1$ and $\cos\theta_0 \simeq \sqrt{2\pi n} \delta +\eps/2$ (with $\eps= \mbr^2/\vew^2$ and $\delta = \mbr/\Lambda$ as defined before). See Appendix~\ref{app:detail_pi/2} for detailed expression for $\cos\theta_0$.

Instead of using Eq.~\eqref{eq:hphi2param} directly, it can be more practical to rewrite the general expression, Eq.~\eqref{eq:hphi2func} as a function of the relaxion-Higgs mixing angle and the relaxion mass
\bea
{\cal L}_{h \phi^2} 
&\simeq&
- \frac{1}{\sqrt 2\vew}\left( 
m_{\phi0}^2 + \frac{1}{2} m_h^2 \sin^2 \theta_{h \phi}
\right) \, h \phi^2 \\
&\simeq& 
- \frac{1}{\sqrt 2\vew}\left( 
m_{\phi}^2 +  \frac{3}{2} m_h^2 \sin^2 \theta_{h \phi}
\right) \, h \phi^2\,.\label{eq:hphi2fin}
\eea
As a result of dynamical relaxion mass tuning, the second contribution in the brackets of Eq.~\eqref{eq:hphi2fin} can significantly exceed the first one -- this feature was extensively analysed in Sec.~\ref{sec:low_obs}. This result contrasts with the case of the untuned relaxion mass, in which $m_{\phi}^2 \simeq m_{\phi0}^2 \gtrsim \vew^2 \sin^2 \theta_{h \phi}$, and therefore
\bea
{\cal L}_{h \phi^2}^{\rm natural} 
\simeq - \frac{m_{\phi}^2}{\sqrt 2\vew} \,h \phi^2\,.
\eea
This natural mass-coupling relation is violated by \eqref{eq:hphi2fin} under the same condition~(\ref{eq:tuning}) as in the case of the mixing angle-mass relation.

\subsection{Experimental probes above eV scale}\label{sec:pheno}

\begin{figure*}
	\centering
	\hspace*{-0.8cm}
	\includegraphics[scale=0.5]{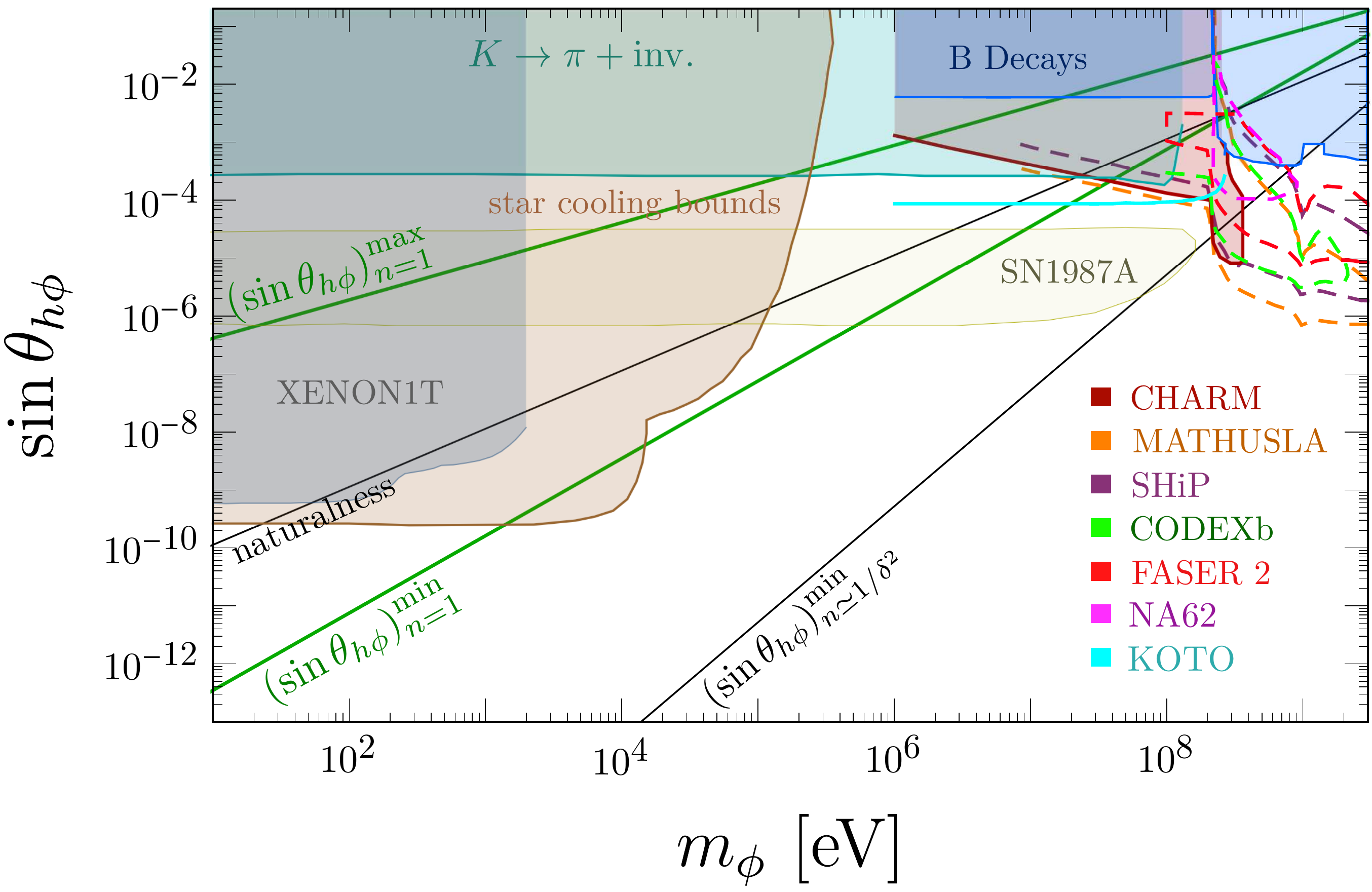}
	\caption{ A relaxion window and the available parameter space for heavier relaxion. 
		The region between the green lines represents the parameter space for relaxion when it is stopped at the first minimum $n=1$. The upper black line represents $\sin\theta_{h\phi} \simeq m_{\phi}/\vew$ denoted as ``naturalness line'' in the figure. The lower black line represents minimum mixing angle for a generic Higgs portal model. The yellow shaded region is constrained by SN1987A~\cite{Turner:1987by,Frieman:1987ui,PhysRevD.39.1020} observation. The brown shaded region is excluded by various star-cooling bounds coming from red giants and horizontal branch stars~\cite{Grifols:1988fv,Raffelt:2012sp,Hardy:2016kme}, where the gray shaded region is excluded by XENON1T~\cite{Budnik:2019olh} experiment for relaxion production at the solar core. Turquoise shaded region is constrained by E949 experiment~\cite{Artamonov:2009sz}. The blue shaded region is excluded by Belle~\cite{Wei:2009zv}, BABAR~\cite{Lees:2013kla} and LHCb~\cite{Aaij:2015tna,Aaij:2016qsm}. The red shaded region is excluded by CHARM~\cite{BERGSMA1985458} experiment, whereas the reach of various future accelerator experiments are plotted in dashed lines. Reinterpretation of recent KOTO result~\cite{KOTOslides,Kitahara:2019lws} is denoted by solid cyan line.       
	}
	\label{fig:pheno}
\end{figure*}

Since the mass of relaxion can vary widely from sub-eV range to a few GeV, we present relevant experimental probes for a relatively heavy relaxion (above eV scale) in this section, while leaving the discussion of a light relaxion to the next section.
In Fig.~\ref{fig:pheno}, we summarize experimental constraints for the relaxion mass  interval $[ 10\eV ,\, 2 \GeV]$.
The constraints dominantly come from the observation of stellar evolution, beam dump experiments, and collider searches. 
For eV -- keV mass range, the scalar couplings to nucleon, electron, and photon are strongly constrained by stellar evolution consideration~\cite{Grifols:1988fv, Cadamuro:2011fd, Raffelt:2012sp, Hardy:2016kme} as those couplings provide alternative channels to stellar energy loss processes. 
For instance, the stringent bound on scalar-electron Yukawa coupling, ${\cal L} \supset - g_{\phi ee}\phi \bar{e}e$, is obtained from the evolution of red giants, constraining the Yukawa coupling as $g_{\phi ee} < 7 \times 10^{-16}$~\cite{Hardy:2016kme}. 
This can be translated as a bound on relaxion-Higgs mixing angle, and is shown as a brown shaded region in the figure.
In addition to the stellar evolution constraints, the relaxion with the mass below keV scale can be copiously produced from the Sun, whose flux can be probed by terrestrial dark matter detectors.
It is shown in~\cite{Budnik:2019olh} that the liquid xenon detectors, such as XENON1T and LUX, place constraints as $g_{\phi ee} \lesssim 2\times 10^{-15}$, which is a factor three weaker than stellar cooling constraints.
This is shown as a gray shaded region in the figure.
We also show the constraint coming from SN1987A~\cite{Turner:1987by, Frieman:1987ui, PhysRevD.39.1020, Krnjaic:2015mbs} as yellow shaded region, while we note the recent critical investigation of SN1987A constraint on light new physics~\cite{Bar:2019ifz}.

For relaxion mass in the range of MeV -- GeV, it can be probed at the luminosity frontier by various beam dump, and accelerator experiments. 
In these experiments, relaxion is dominantly produced in rare decays of $K$ and $B$-meson.
The CHARM experiment performed a search for axion-like particles decaying to $e^{+}e^{-}$, $\mu^{+}\mu^{-}$ and/or $\gamma\gamma$~\cite{BERGSMA1985458}. 
Their result on axion-like particle was recasted to constrain Higgs-portal models in~\cite{Clarke:2013aya}, which is shown as red shaded region in the figure. 
In addition, $B$-meson decays can be constrained from Belle~\cite{Wei:2009zv}, BaBar~\cite{Lees:2013kla}, and LHCb~\cite{Aaij:2016qsm,Aaij:2015tna}. 
Based on $B^\pm \to K^\pm \phi \to K^\pm \ell^+ \ell^-$ ($\ell = e,\mu$), and $B \to K + {\rm invisible}$, the relaxion-Higgs mixing angle is constrained as in the blue shaded region in the figure~\cite{Clarke:2013aya}. 
For relaxions $100\keV <m_{\phi}\leq 10\MeV$, E949 experiment provides a stringent constraint on relaxion-Higgs mixing angle from $K\to \pi +{\rm inv.}$~\cite{Artamonov:2009sz}, which is shown as turquoise shaded region in the figure.
A bound from invisible $K$ decay could be potentially improved by NA62 experiment in the beam dump mode~\cite{Flacke:2016szy} (see also~\cite{Moulson:2702688,Beacham:2019nyx}). Recent KOTO result of $K_L\to \pi^{0}\nu\bar{\nu}$~\cite{KOTOslides} has been reinterpreted as $K\to \pi+\rm{NP}(\phi)$ searches~\cite{Kitahara:2019lws} and the bound is represented as the solid cyan line.
The sensitivity reach of future accelerator experiments like MATHUSLA~\cite{Curtin:2018mvb,Evans:2017lvd}, SHiP~\cite{Strategy:2019vxc,Lanfranchi:2243034}, CODEX-b~\cite{Beacham:2019nyx, Gligorov:2017nwh}, and FASER-2~\cite{PhysRevD.97.055034,Beacham:2019nyx} are also shown, and denoted by dashed orange, violet, green, and red lines respectively in Fig.~\ref{fig:pheno}. 
See~\cite{Flacke:2016szy, Frugiuele:2018coc} for a more detailed discussion of these constraints. 
Although the relaxion Higgs cross quartic coupling is enhanced as discussed in Section~\ref{sec:xquartic}, we have taken $\lambda_{H\phi}\to 0$ limit while projecting these bounds to relaxion parameter space.

\subsection{Experimental probes below eV scale}\label{sec:pheno_low}
The relaxion below eV scale can mediate a long-range force between matter particles. 
For mass below eV scale, so-called fifth force experiments as well as equivalence principle tests provide a powerful way to probe the existence of a light scalar field.
These experiments constrain a large fraction of available parameter space below eV scale~\cite{PhysRevD.32.3084, BORDAG20011, Kapner:2006si, Schlamminger:2007ht, Bordag:2009zzd, PhysRevLett.120.141101}.

Additional probe is available if the relaxion accounts for the dark matter relic density in the present universe.
The possibility of relaxion being dark matter was briefly mentioned in the original paper~\cite{Graham:2015cka}, and it is shown in~\cite{Banerjee:2018xmn} that the relaxion could be coherently oscillating dark matter for $10^{-11}\eV \lesssim m_\phi \lesssim 10^{-8}\eV$ if the reheating temperature is higher than the critical temperature of the EW phase transition. 
To investigate the phenomenological consequence of relaxion dark matter, we write the low energy effective Lagrangian of the relaxion,
\bea
\!\!\! {\cal L}_{\rm eff} \supset 
- \frac{1}{4} FF
- \frac{1}{4} GG
- \sin\theta_{h\phi} \frac{\phi}{\sqrt 2}
\left( 
\sum_{f} \frac{m_f}{\vew}  \bar{f} f 
+ c_{\gamma} \frac{\alpha}{4\pi \vew} FF 
+ c_{g} \frac{\alpha_s}{4\pi \vew} GG \right),
\eea
where $c_{\gamma}$ and $c_{g}$ are an order one coefficient. 
Nonvanishing background field value $\langle \phi \rangle = \sqrt{2 \rho_{\rm DM}} / m_\phi \cos(m_{\phi}t)$, where $\rho_{\rm DM} = 0.4 \GeV/{\rm cm}^3$ is the local dark matter density, induces a small oscillating component for the mass of electron and nucleon, and also for electromagnetic and strong coupling constants, 
\bea
\frac{\Delta m_f}{m_f} &\simeq&  \sin\theta_{h\phi} \frac{\phi}{\sqrt{2}\vew}\, ,
\label{v1}
\\
\frac{\Delta \alpha}{\alpha} &\simeq& - c_{\gamma} \sin\theta_{h\phi} \frac{\alpha \phi}{\sqrt{2} \pi \vew}\,,
\\
\frac{\Delta \alpha_s}{\alpha_s} &\simeq& - c_{g} \sin\theta_{h\phi} \frac{\alpha_s \phi}{\sqrt{2} \pi \vew}\, .
\label{v3}
\eea
The atomic clock transitions can be used to probe such oscillations of fundamental constants~\cite{Arvanitaki:2014faa}.
 
We briefly discuss the projected sensitivity from a clock comparison test, while we refer interested readers to~\cite{Arvanitaki:2014faa, Safronova:2017xyt} for more detailed descriptions. 
Consider an atom $A$ and $B$ and corresponding clock transition frequencies $f_{A,B}$. 
In general, each of clock transition frequencies can be written as a function of fundamental constants, 
\bea
f_i = h_i( \alpha, \alpha_s , m_e, m_q),
\eea
where $h_i$ is some function of fundamental constants, and $i=A,\,B$. 
Since all fundamental constants are oscillating due to the background dark matter, the ratio of clock frequencies $f_A/f_B$ will oscillate in time with a frequency equal to the mass of dark matter.
The fractional change of this observable can be written as
\bea
\frac{\Delta(f_A/f_B)}{f_A/f_B} 
=
\sum_{y = \{ \alpha,\alpha_s,m_e,m_q\}} \left( \frac{\partial \ln h_A}{\partial \ln y } - \frac{\partial \ln h_B}{\partial \ln y} \right) \frac{\Delta y}{y} \, .
\label{ff}
\eea
The quantities in the parenthesis, $\partial \ln h_A / \partial \ln y$, are referred to as sensitivity coefficients, and are in general different for different clock systems (see e.g.~\cite{Arvanitaki:2014faa, Safronova:2017xyt} for a list of sensitivity coefficients for various clocks). 
This fractional change of the transition frequency is measured after averaging over time $\tau$, and the measurement is repeated until the total experimental time scale $T$ is reached. 
This procedure constitutes a time series of $\Delta(f_A/f_B)/(f_A/f_B)$, and discrete Fourier transform allows one to find whether there is an excess power at a certain frequency due to the dark matter.

\begin{figure}
\centering
\includegraphics[scale=0.5]{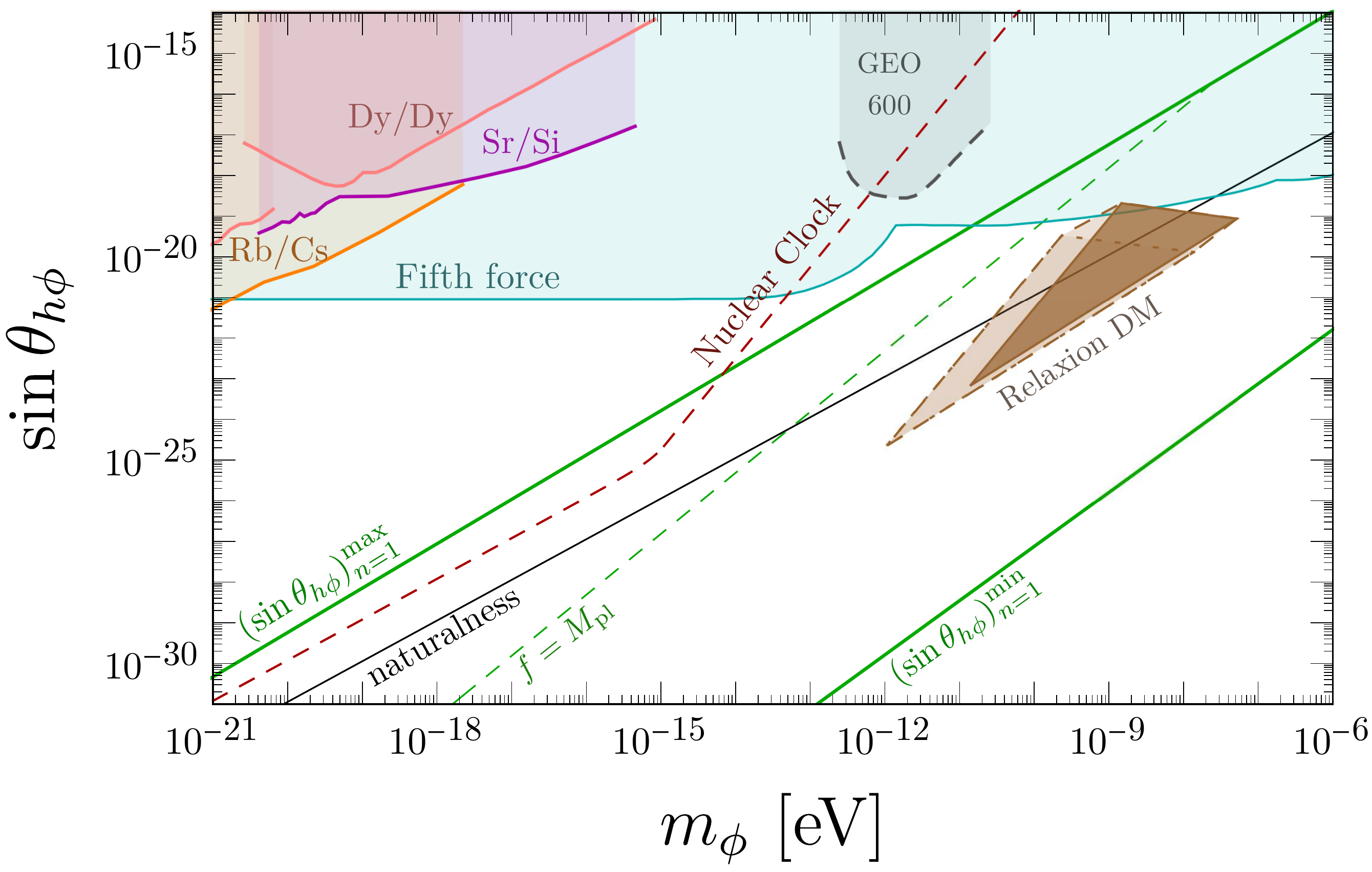}
\caption{A relaxion window and the available parameter space for the light relaxion. The green lines are the maximum and minimum mixing angle at $n=1$, while the black line corresponds to the naturalness line, $\sin\theta_{h\phi}\sim m_\phi /\vew$. 
If one requires $f$ to be sub-Planckian, the mixing angle should be smaller than the green dashed line. 
The fifth force constraints~\cite{Schlamminger:2007ht, PhysRevLett.120.141101}, represented as turquoise shaded region, provide stringent constraints over the mass scale shown in the figure.
The relaxion constitutes the entire dark matter in the universe in the brown shaded region, while it only constitutes $>1\%$ of the dark matter in the light shaded region. 
This DM parameter space is obtained by projecting the relaxion dark matter parameter space for all available $\Lambda$ onto this two dimensional parameter space. 
The existing atomic clock bounds, using Dysprosium (pink)~\cite{PhysRevLett.115.011802}, Rubidium-Cesium (orange)~\cite{PhysRevLett.117.061301}, and Strontium-Silicon cavity (magenta)~\cite{junye:private}, are also shown.  
Interferometry can also be used to probe dark matter, and the projected sensitivity of GEO 600 experiment is shown as gray dashed line~\cite{PhysRevResearch.1.033187}. 
Finally, we show the projected sensitivity of nuclear clock as a red dashed line. 
See the main text for details. }
\label{fig:low_mass}
\end{figure}

The clock stability is described by Allan deviation $\sigma_y(\tau) \propto 1 / \sqrt{\tau}$.
The signal-to-noise ratio is
\bea
{\rm S/N} \sim \frac{\big[ \Delta (f_A /f_B)/(f_A/f_B) \big]_{\tau}}{\sigma_y(\tau)} F(\tau_{\rm int})
\label{snr}
\eea
where $\tau_{\rm int}$ is total integration time. 
The function $F$ is defined as
\bea
F(\tau_{\rm int}) =
\begin{cases}
 \sqrt{\tau_{\rm int}} \sqrt{\rm Hz} & \textrm{for } \tau_{\rm int}<\tau_{\rm coh}
 \\
  (\tau_{\rm int} \tau_{\rm coh})^{1/4} \sqrt{\rm Hz} & \textrm{for } \tau_{\rm int}>\tau_{\rm coh}\, ,
 \end{cases}
\eea
where $\tau_{\rm coh} = 2\pi/mv^2 $ is dark matter coherence time with the virial velocity $v \sim 10^{-3}$.  
The numerator of Eq.~\eqref{snr} should be understood as an amplitude of the fractional frequency change averaged over $\tau$. 

In Fig.~\ref{fig:low_mass}, we project a reach of future nuclear clock transition. 
We assume that the clock instability is dominated by the quantum projection noise (QPN) of nuclear clock, and obtain the projected sensitivity by solving ${\rm S/N}=1$. A particularly simple expression for QPN-limited $\sigma_y(\tau)$ is available if the nuclear clock transition is probed by Ramsey method. In this case, one finds~\cite{PhysRevA.47.3554}
\bea
\sigma_y(\tau) = \frac{1}{\omega_0} \frac{1}{\sqrt{T \tau}} \, ,
\eea
where $\omega_0 \simeq 8.3\eV $ is the energy of nuclear clock transition of $^{229}$Th~\cite{Seiferle:2019fbe}, $T$ is the time interval between two $\pi/2$-pulses in Ramsey method, and $\tau$ is averaging time (see also~\cite{RevModPhys.87.637} for a review). 
On the other hand, the fractional change of the frequencies is given as
\bea
\frac{\Delta f_A/f_B}{f_A/f_B} \simeq 10^4 \frac{\Delta \alpha}{\alpha} + 10^5 \frac{\Delta (m_q / \Lambda_{\rm QCD})}{(m_q/\Lambda_{\rm QCD})}
\sim 10^5 \frac{\phi}{\vew} \sin\theta_{h\phi}\,,
\label{frac_freq}
\eea
where $m_q$ is the light quark mass, and $\Lambda_{\rm QCD}$ is the QCD scale which is directly related to $\alpha_s$ via dimensional transmutation~\cite{Flambaum:2006ak,Berengut:2010zj}. Here, the subscript $A$ denotes nuclear clock, while the subscript $B$ could be an optical lattice clock system. 
For the projected sensitivity, we choose the averaging time $\tau = 1$ sec, $T=0.5$ sec, and $\tau_{\rm int} = 10^6$ sec.
The result is shown as a red dashed line in Fig.~\ref{fig:low_mass}. 
For the relaxion mass $m_\phi > 2\pi / \tau$, the fractional frequency change oscillates many times, and thus, averaged value of Eq.~\eqref{frac_freq} has additional $1/m_\phi$ suppression~\cite{Derevianko:2016vpm}. 
We note that by the sensitivity to a specific region of the parameter space could be improved in principle, upon the usage of dynamic decoupling at high frequencies, or a longer averaging time and/or adding more atoms.

We also summarize existing bounds for light relaxion. 
Fifth force experiments provide a stringent constraint over a wide range of masses without an assumption of relaxion being dark matter~\cite{Schlamminger:2007ht, PhysRevLett.120.141101}. 
For masses below $m_\phi = 10^{-15}\eV$, atomic clocks, for instance, isotopes of Dysprosium (pink)~\cite{PhysRevLett.115.011802}, Rubidium-Cesium atomic fountain (orange)~\cite{PhysRevLett.117.061301}, and Strontium-Silicon cavity (purple)~\cite{junye:private}, constrain some fraction of parameter space. 
In addition, interferometer can also be used to probe relaxion dark matter as the temporal oscillation of DM can change the optical path length of each arm of interferometry~\cite{PhysRevResearch.1.033187}. 
We also show the available parameter space of relaxion dark matter studied in~\cite{Banerjee:2018xmn}. 
In the minimal scenario, the relaxion dark matter can be realized for the cutoff scale, $\Lambda_{\rm min} < \Lambda \lesssim {\cal O}(10^2) \TeV$.
The brown shaded region in the figure is the parameter space where the relaxion  constitutes entire dark matter in the present universe, while the light shaded region is where the relaxion constitutes more than one percent of total dark matter. 
Note that the above DM parameter space is the result of the projection of the available parameter space for all possible $\Lambda$ onto two-dimensional parameter space plane $(m_\phi,\,\sin\theta_{h\phi})$. This is contrary to what was shown in~\cite{Banerjee:2018xmn, Banerjee:2019epw} where a particular cutoff is chosen.

We note that the clock comparison test is a specific example to probe the oscillation of fundamental constants. In general, any frequency comparison with different dependence on the constants of nature would serve the purpose. For example atom-cavity comparison can be used, which along with some recent developments, is discussed in~\cite{PhysRevLett.123.141102, Aharony:2019iad,Stadnik:2015xbn}. (see also~\cite{Antypas:2019yvv} for a recent discussion.)

\begin{figure}
	\centering
	\includegraphics[scale=0.21]{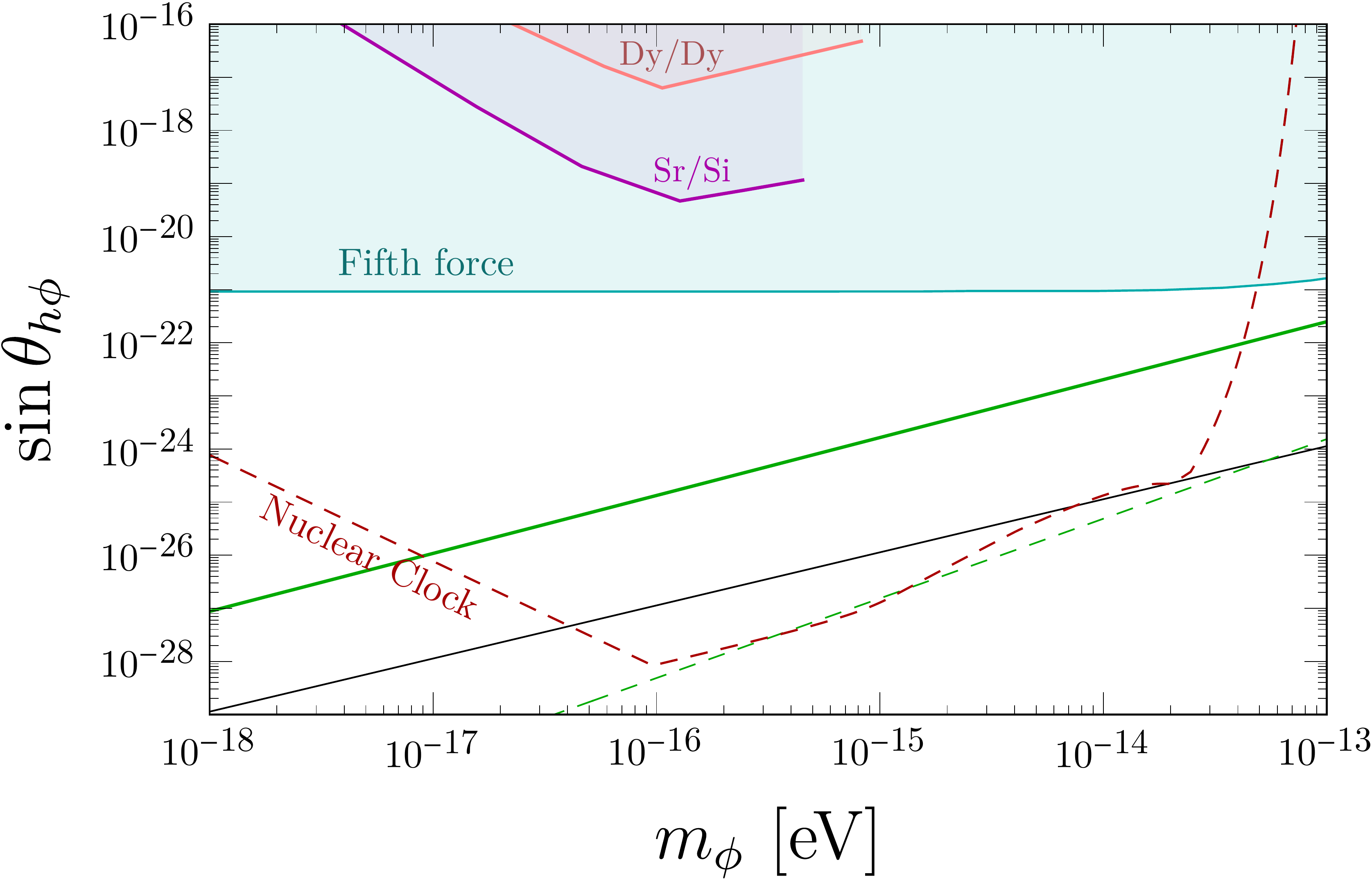}
	\,
	\includegraphics[scale=0.21]{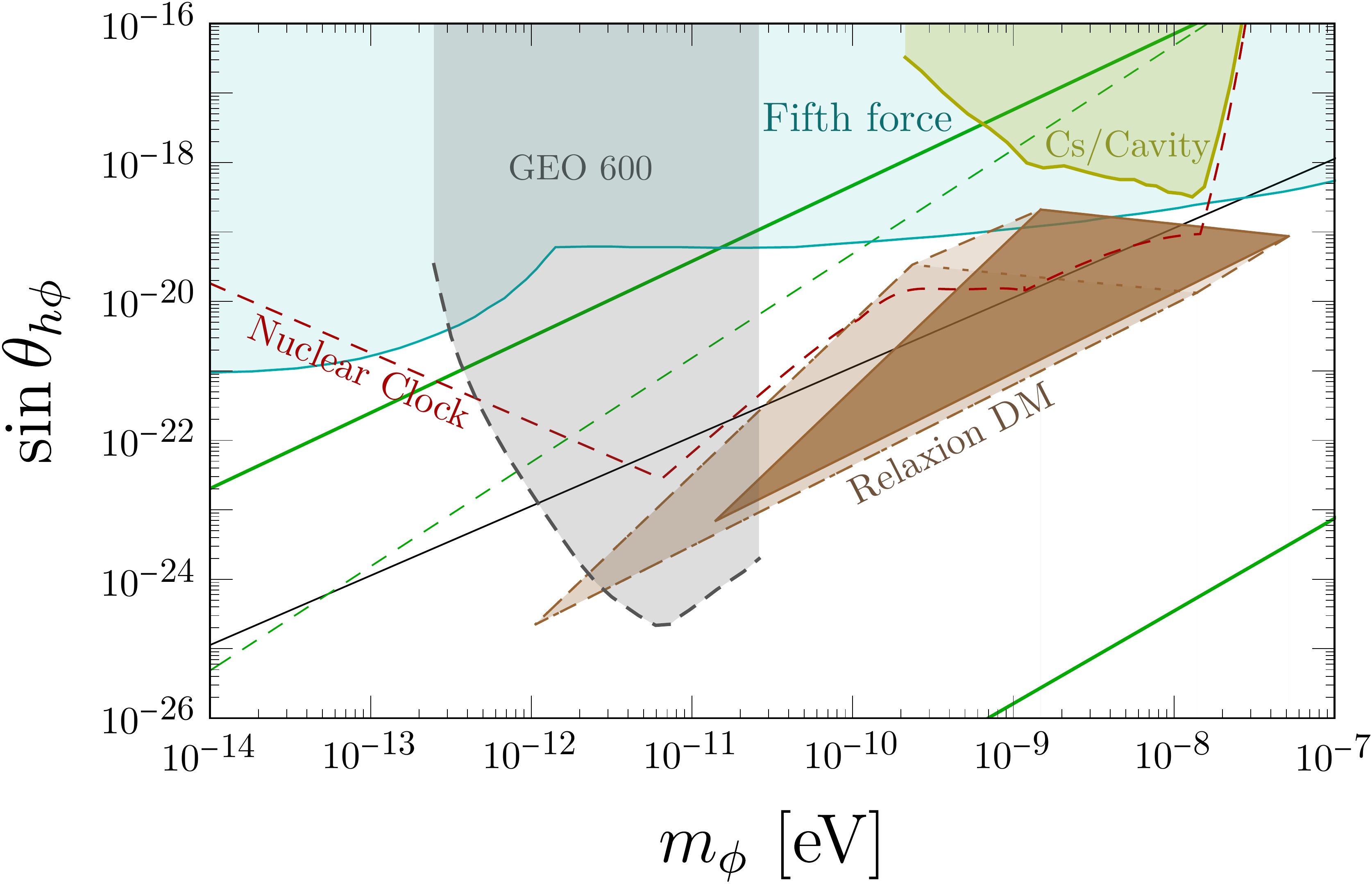}
	\caption{A relaxion window and updated parameter space for the light relaxion in the presence of a solar halo (left), and an earth halo (right). In the presence of such halos, the sensitivity of the various clock tests described in Fig.~\ref{fig:low_mass} enhances because of the density and longer coherence time of those objects as described in~\cite{Banerjee:2019epw,Banerjee:2019xuy}. (Right) The yellow shaded region describes the sensitivity reach of Cesium clock-cavity comparison test~\cite{PhysRevLett.123.141102} in the presence of an earth halo.		
					}
	\label{fig:halo}
\end{figure}

We also note that, if a compact boson star consisting of $\phi$ forms in the early universe (see e.g.~\cite{Arvanitaki:2019rax,Fonseca:2019ypl}), and is gravitationally bounded to stars or even planets, the projected sensitivity can be greatly enhanced since the effect is proportional to the square root of background density and enjoys potentially a much longer coherent time, as discussed in~\cite{Banerjee:2019epw, Banerjee:2019xuy}. This is shown in Fig.~\ref{fig:halo}. 

\section{Discussion}\label{sec:conclusion}
We have examined the dynamics of cosmological relaxion around the local minima.
We have found that generically the relaxion is stabilized at the shallow part of the potential, suppressing the relaxion mass by a small parameter $\delta = \mbr/\Lambda$, i.e. not only the Higgs mass but also the relaxion mass is relaxed due to the dynamical relaxation mechanism.

The parametric suppression of relaxion mass leads to interesting consequences regarding low energy phenomenology.
Due to the relaxation of the relaxion mass, the resulting mixing angle at a given mass is larger than what is required for the radiatively stable scalar mass, $\sin\theta_{h\phi} = m_\phi/\vew$. In other words, once a light scalar is found, low energy observer might consider the observed value of mass and the mixing angle unnatural in view of conventional naturalness argument, despite all of underlying model parameters are technically natural.
We have also found that the maximum mixing angle could be at most three orders of magnitude larger than $\sin\theta_{h\phi} = m_\phi /\vew$ when the relaxion mass is around eV scale. This, in turn, may result in the corresponding enhancement of the signals in various experiments searching for light scalar fields.

\begin{figure}
	\centering
	\includegraphics[scale=0.39]{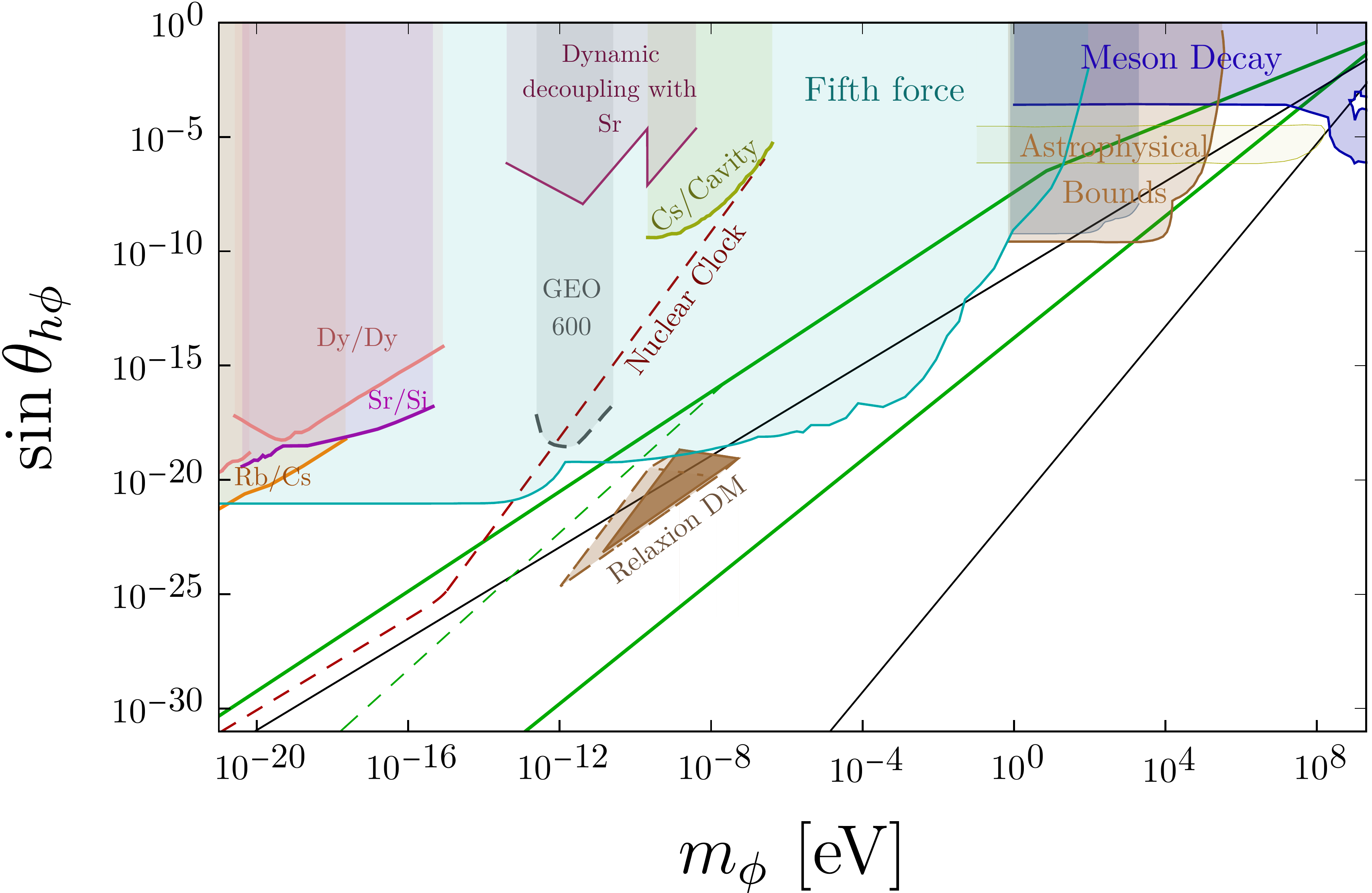}
	\caption{
		Updated parameter space for relaxion. The region between two solid green lines denotes the parameter space for relaxion when it stops at the first minimum. 
		The region between the black solid lines represents the parameter space for relaxion when it stops at a generic minima (see the discussion in Sec.~\ref{sec:genmix}). The region above the dashed green line represents super-Planckian decay constant. The brown triangular region represents relaxion DM parameter space as discussed in~\cite{Banerjee:2018xmn}.
		The blue, light yellow, light brown, and the light black shaded regions on the top right corner describe excluded parameter space from various collider  experiments and astrophysical considerations. These are discussed in more detail in Section~\ref{sec:pheno} and in Fig.~\ref{fig:pheno}. 
		The turquoise, light orange, magenta, pink, and grey dashed shaded region represents constraints on sub-$\eV$ relaxion scenario from various fifth force and clock-comparison experiments which has been discussed in  Section~\ref{sec:pheno_low} and in Fig.~\ref{fig:low_mass}. The purple shaded region is excluded by recent clock caparison test with dynamic decoupling~\cite{Aharony:2019iad}, while the darker yellow shaded region is excluded by Cesium clock-cavity comparison test~\cite{PhysRevLett.123.141102}.      
		 }
	\label{fig:overview}
\end{figure}

We have also updated the relaxion parameter space, which is represented in Fig.~\ref{fig:overview}.  
We have seen that the constraints from colliders and beam dump experiments already excluded the possibility that the relaxion is stabilized at $n=1$ if its mass is above ${\cal O}(100)\MeV$. 
In addition, star cooling bounds probed a significant fraction of available parameter space for $\keV < m_\phi < 100\keV$, while fifth force experiments and the equivalence principle tests provide stringent constraint below eV scale.
Furthermore, we have discussed additional probes when the relaxion constitutes dark matter in present universe.
Because of the dependence on the relaxion field value,  the fundamental constants have an oscillating component, which can be efficiently probed by atomic clocks. 
We have briefly discussed the reach of future nuclear clock transition. 

\acknowledgments
We would like to thank Nitzan Akerman, David Leibrandt, David Hume, Yevgeny V. Stadnik, and Jun Ye for useful discussions. We would also like to thank Diego Redigolo and Lorenzo Ubaldi for initial collaboration on this project. 
The work of OM is supported by the Foreign Postdoctoral Fellowship Program of the Israel Academy of Sciences and Humanities.
The work of GP is supported by grants from The U.S.-Israel Binational
Science Foundation (BSF), European Research Council (ERC), Israel
Science Foundation (ISF), Yeda-Sela-SABRA-WRC, and the Segre Research Award. 
The work of MS is part of the Thorium Nuclear Clock project that has received funding from the European Research Council (ERC) under the European Union's Horizon 2020 research and innovation programme (Grant agreement No. 856415).

\appendix
\section{Detailed analysis of relaxion stopping point}\label{app:detail_pi/2}

We consider a relaxion field space where the Higgs vacuum expectation value is close to the electroweak scale.
In this case, the curvature of the potential along the Higgs direction is larger than that along the relaxion direction, $|\partial^2 V / \partial H^2| \gg |\partial^2 V / \partial \phi^2|$, and this allows us to assume that the Higgs adiabatically follows its instantaneous vacuum.
At the minimum along the Higgs direction, the relaxion effective potential is
\bea
V_{\rm eff}(\phi) = - g \Lambda^3 \phi -  [v^2(\phi) ]^2\,.
\eea
As in the main text, we write the field variable as
\bea
\frac{\phi}{f} = 2 \pi m + \theta\,,
\eea
with $m \in {\mathbb Z}$ and $\theta \in [ 0 , 2\pi)$. 
The relaxion-dependent Higgs vacuum expectation value is
\bea
v_m^2(\theta)= \frac{1}{2} \left( - \Lambda^2 + \frac{\Lbr^4}{\Lambda^2} (2\pi m + \theta) + \mbr^2 \cos \theta \right).
\eea
We have taken Higgs quartic as $\lambda= 1$ for simplicity. 
With the effective description of relaxion potential, the relaxion classically stops for the first time when $V_{\rm eff}'=0$ is satisfied, which leads to the following condition, 
\bea
\sin \theta = \frac{\vew^2}{v^2_m(\theta)} + \frac{\vew^2}{\Lambda^2}\,.
\label{master}
\eea
Before the Higgs VEV reaches $\vew$, the right hand side of the equation is always larger than unity, and thus, no solution exists. 
On the other hand, as we have already discussed in the main text, the Higgs VEV changes incrementally, $\Delta v^2 /v^2 \simeq \delta^2$.
Therefore, the first solution starts to appear when the right hand side is close to unity up to ${\cal O} (\delta)$ factor, meaning that local minimum and maximum should appear close to $\theta \simeq \pi/2$. 

After rearranging the Higgs VEV, we find
\bea
v^2_m(\theta) = \vew^2 \left( - \frac{\Lambda^2}{2\vew^2} + \pi m \delta^2 +  \frac{\delta^2}{2}\theta + \frac{\mbr^2}{2\vew^2} \cos\theta \right).
\eea
Note that all terms in the parenthesis are small but $\Lambda^2 / \vew^2$, which originates from UV sensitive squared Higgs mass term.
This should be canceled by the relaxion evolution, and in the above description, we can do it by shifting $m$. 
To study the field space around $\langle H \rangle \sim \vew$, we shift $m \to m + \frac{1}{\pi \delta^2}(1 +  \frac{\Lambda^2}{2\vew^2})$, and find
\bea
v_m^2 (\theta) =  \vew^2 \left( 1 +  \pi (m + \alpha) \delta^2 + \frac{1}{2} \delta^2 \theta + \frac{\mbr^2}{2\vew^2} \cos \theta \right)\,,
\eea
where a fuzzy factor $\alpha \in [ 0 ,1)$ is introduced since the shift $(\pi\delta^2)^{-1} (1+\frac{\Lambda^2}{2\vew^2})$ need not be an integer.

Equipped with the expression of the Higgs VEV adjacent to the electroweak scale, we approximate the master equation Eq~.\eqref{master} as
\bea
\sin\theta 
\simeq 1 - \pi(m+\alpha) \delta^2 - \frac{1}{2} \delta^2 \theta - \frac{\mbr^2}{2\vew^2} \cos\theta + \frac{\vew^2}{\Lambda^2}\,.
\eea
Again, the solution would appear around $\theta \sim \pi/2$, we expand trigonometric function around $\pi/2$, and find
\bea
(\pi/2 - \theta) \simeq \frac{1}{2} \frac{\mbr^2}{\vew^2}  \pm \sqrt{2\left(\pi(\alpha+m)\delta^2 + \frac{\pi\delta^2}{4} - \frac{\vew^2}{\Lambda^2} \right) + \left( \frac{\mbr^2}{2\vew^2} - \frac{\delta^2}{2} \right)^2}\,.
\eea
Terms other than $\pi m \delta^2$ inside the parenthesis would not matter for determining the local minimum and maximum since in any cases, the first solution appears when the whole term in parenthesis becomes positive. 
Thus, we find
\bea
\theta = \left(\frac{\pi}{2} - \frac{1}{2} \frac{\mbr^2}{\vew^2} \right) \pm \delta \sqrt{2\pi \alpha}\,\,,
\eea
with a fuzzy factor $0\leq \alpha <1$. 
The local minimum and maximum are centered around $\pi/2 - \mbr^2/ 2 \vew^2$, and separated with each other by $\sim 2 \delta$.
For $n$-th local minimum, we can easily shift $m \to m+1$, and the resulting separation would be $\sqrt{n}\delta$.

\bibliographystyle{JHEP}
\bibliography{ref}
\end{document}